\documentclass[aip,jcp,reprint]{revtex4-1}

\usepackage{graphicx}
\usepackage{dcolumn}
\usepackage{bm}
\usepackage{amsmath}
\usepackage{amssymb}
\usepackage{mathrsfs}
\usepackage{lmodern}
\usepackage{mhchem}
\usepackage[cspex,bbgreekl]{mathbbol}
\usepackage{color}
\usepackage{stackrel}

\newcommand{\bea}{\begin{eqnarray}}
\newcommand{\eea}{\end{eqnarray}}

\newcommand{\be}{\begin{equation}}
\newcommand{\ee}{\end{equation}}
\newcommand{\ba}{\begin{eqnarray}}
\newcommand{\ea}{\end{eqnarray}}

\begin{document}

\title{Reaction kinetics in open reactors and serial transfers between closed reactors}
\author{Alex Blokhuis}
\affiliation{Laboratoire de Physico-Chimie Th\'eorique - UMR CNRS Gulliver 7083,\\ PSL Research University, ESPCI, 10 rue Vauquelin, F-75231 Paris, France}
\affiliation{Laboratory of Biochemistry, PSL Research University,
ESPCI, \\ 10 rue Vauquelin, 75231 Paris Cedex 05, France}
\author{David Lacoste}
\affiliation{Laboratoire de Physico-Chimie Th\'eorique - UMR CNRS Gulliver 7083,\\ PSL Research University, ESPCI, 10 rue Vauquelin, F-75231 Paris, France}
\author{Pierre Gaspard}
\affiliation{Center for Nonlinear Phenomena and Complex Systems, Universit\'e Libre de Bruxelles (U.L.B.), Campus Plaine, Code Postal 231, B-1050 Brussels, Belgium}

\date{\today}

\begin{abstract}
Kinetic theory and thermodynamics of reaction networks are extended to the out-of-equilibrium dynamics 
of continuous-flow stirred tank reactors (CSTR) and serial transfers.  On the basis of their stoichiometry matrix, the conservation laws and the cycles of the network are determined for both dynamics.  It is shown that the CSTR 
and serial transfer dynamics are equivalent in the limit where the time interval between the transfers tends to
 zero proportionally to the ratio of the fractions of fresh to transferred solutions. These results are illustrated
 with a finite cross-catalytic reaction network and an infinite reaction network describing mass
 exchange between polymers. Serial transfer dynamics is typically used in molecular evolution experiments in the context of research on the origins of life. The present study is shedding a new light on the role played by serial transfer parameters in these experiments.
\end{abstract}

\pacs{
05.70.Ln,  
05.70.-a,  
82.20.-w   
}

\maketitle

\section{Introduction}

The regulation of self-assembly plays a critical role in biological systems, 
both for the emergence of life out of non-living matter 
and for its maintenance. Remarkable advances
in the manipulation, replication, and sorting of information-rich biopolymers, 
such as nucleic acids \cite{Vaidya2012} or peptides,\cite{Forsythe2017}
allow us to perform novel proofs of principle regarding the mechanisms prevailing to the emergence of life.  In this field, chemical reaction networks of interdependent molecular species have long been considered as a central element for theoretical studies and simulations.\cite{Eigen1971,EP77,EP78a,EP78b,Eigen1992,Kauffman1993,Segre2000}
Thanks to the aforementioned experimental advances,\cite{Vaidya2012,Forsythe2017}
these systems are now accessible to experiments.\cite{Lincoln2009,Matsumura2016} 
Besides the relevance for the origins of life, such molecular evolution experiments
suggest new chemical pathways to achieve the self-assembly 
of molecular elements into complex molecules and beyond into supramolecular 
structures.\cite{Zwaag2015,Zeravcic2017}
Directed evolution experiments are a special kind of molecular evolution experiment, 
in which variations due to mutations are introduced artificially while a well-controlled selection  
pressure is applied.\cite{Agresti2010}
This allows one to select enzymes with an improved efficiency, which is 
particularly attractive for industrial applications in biotechnology.\cite{Arnold1997}

A common feature in these molecular evolution experiments is that 
they are open reaction networks, which are maintained  
out of equilibrium through incoming and outgoing fluxes of 
molecules or energy.
Many examples of such structures exist in biology. 
Cytoskeletal filaments such as actin or microtubules 
display a rich dynamics which can only exist 
through a constant flux of ATP or GTP hydrolysis.\cite{Hill1989,Ranjith2012,Jegou2016}
The same is true for larger cellular structures 
such as membrane protein clusters or P-granules,\cite{Brangwynne2009,Zwicker2017} which owe 
their special liquid-like properties to the turnover of their constituents. 
These systems fall into the broad class of active systems, which are presently the focus 
of intense research both in physics and biology.\cite{Marchetti2013}
Active systems typically form dissipative structures which would 
not exist in the absence of non-equilibrium fluxes from the environment and which manifest
very different properties than at equilibrium.\cite{NP77} Therefore, an approach based on non-equilibrium statistical mechanics and thermodynamics is required to describe them.
Building on the well established framework of nonequilibrium thermodynamics \cite{P67,GM84} 
and on more recent progress in stochastic thermodynamics, 
a generic and comprehensive theory of open chemical networks 
has been recently developed.\cite{Polettini2014,RE16}
In previous work, this theoretical framework has been used to analyze a mass-exchange 
model of polymers with identical monomers in a closed system \cite{Lahiri2015} 
and an open version of the same model, in which chemostats fix the concentrations 
of polymers of certain length and as a result drive the system out of equilibrium.\cite{Rao2015a} 
When different monomer types are present, an even richer dynamics of recombination 
between polymer chains is possible due to the 
interplay between polymer lengths and polymer sequences.\cite{Blokhuis2017}

There exist several approaches to drive a reaction network out of equilibrium. One of them is the continuous-flow stirred tank reactor (CSTR), in which a well-stirred solution is continuously fed by reactants while keeping constant its volume with a compensating outflow.\cite{A89,BPV84,VL88,N95,EP98}

In the experiment of Ref.~\onlinecite{Vaidya2012}, 
a mixture of interdependent biopolymers evolves through serial transfers, in which 
a part of the solution of interest is periodically transferred to 
a nutrient medium, from which the solution of interest draws reactant molecules and energy.
Chemical systems evolving by serial transfers have similarities with systems evolving in 
CSTRs, but it is not clear whether the two dynamics are completely equivalent from a kinetic or thermodynamic point of view. 
CSTRs can exhibit a large range of dynamic phenomena, such as stationary, oscillatory, 
multi-stable or chaotic \cite{VL88,A89,S91} and it is natural to ask whether all these regimes 
are possible in a reactor evolving instead by serial transfers. 
Let us also mention that a setup somehow similar to CSTRs also exists under the name of chemostats 
in studies of the metabolism of cells: in such bioreactors, a population of cells is
 maintained in an exponentially growing phase by the injection of nutrients into the system.\cite{Salman2012}

In this paper, we compare the kinetic and thermodynamic descriptions 
of open reactors (CSTRs) with that obtained in the case of serial transfers between closed reactors. We illustrate our results with a study of the polymer mass-exchange
model of Ref.~\onlinecite{Rao2015a}, except that now the system is not driven 
out of equilibrium by chemostats as considered in Refs.~\onlinecite{Polettini2014,RE16}, but by  
matter fluxes in a CSTR configuration. 
This paper is organized as follows: 
in Sec.~\ref{sec:CSTR}, we study the kinetics and thermodynamics of CSTRs, which is then 
illustrated with a couple of examples of chemical reactions, then in Sec.~\ref{sec:ST} 
we carry out the corresponding study for the case of serial transfer dynamics between closed reactors.  The conclusion is drawn in Sec.~\ref{conclusion}.

\section{Continuous-flow stirred tank reactor}
\label{sec:CSTR}

\subsection{Kinetic equations of the CSTR}

\begin{figure}[h!]
\includegraphics[scale=0.8]{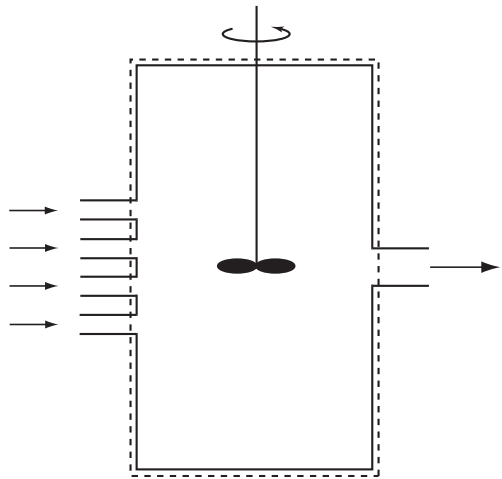}
\caption{Schematic representation of a continuous-flow stirred tank reactor (CSTR).  The dashed line depicts a fictitious surface delimiting the volume $V$ of the reactor.}
\label{fig1}
\end{figure}

Continuous-flow stirred tank reactors are open reactors with a continuous feed of reactants and an outflow in order to keep the volume constant inside the reactor (see Fig.~\ref{fig1}).  
The reactants are pumped into the reactor at given controlled concentrations $c_{k,{\rm in}}$.  
The solution in the reactor is well stirred so that the concentrations of the different species 
can be supposed to remain uniform inside the volume of the reactor.  In order to establish the
 evolution equations of the concentrations in the CSTR, we use the balance equations of the 
concentrations $c_k$ in the flow:
\be
\partial_t c_k + \pmb{\nabla}\cdot\left( c_k{\bf v}+{\bf j}_k\right) = \sum_i \nu_{ki} w_i \, ,
\label{eq-conc}
\ee
expressed in terms of the fluid velocity $\bf v$, the diffusive current density of species $k$ given by Fick's law ${\bf j}_k=-D_k\pmb{\nabla} c_k$, the stoichiometric coefficient $\nu_{ki}$ of species $k$ in the reaction $i$, and the rate $w_i$ of reaction $i$.  The different species are passively advected by the turbulent velocity field $\bf v$ of the flow.  By stirring, the concentrations rapidly become uniform so that the Fickian diffusive current densities are soon negligible ${\bf j}_k\simeq 0$.  
Integrating the balance equation~(\ref{eq-conc}) over the volume $V$ of the reactor, we find
\be
\int_V\partial_t \, c_k \, dV  + \int_{\partial V}c_k{\bf v}\cdot d{\bf A} = \int_V \sum_i \nu_{ki} w_i \, dV \, ,
\label{int-eq-conc}
\ee
where $d{\bf A}$ is the surface element of integration on the border $\partial V$ of the volume $V$.  The surface integral has contributions from the inflow tube of species $k$ entering with concentration $c_{k,{\rm in}}$ and the outflow tube where the species $k$ exits at the uniform concentration $c_k$ resulting from stirring:
\be
 \int_{\partial V}c_k{\bf v}\cdot d{\bf A} =  \int_{\partial V_{k,{\rm in}}}c_k{\bf v}\cdot d{\bf A} + \int_{\partial V_{\rm out}}c_k{\bf v}\cdot d{\bf A}  \, .
\ee
Since the concentrations can be supposed to be uniform at entry and exit, we get
\be
\int_{\partial V}c_k{\bf v}\cdot d{\bf A} =  -\phi_{k,{\rm in}}\, c_{k,{\rm in}} + \phi_{\rm out} \, c_k  
\ee
in terms of the ingoing flux $\phi_{k,{\rm in}} = \int_{\partial V_{k,{\rm in}}} {\bf v}\cdot d{\bf A}$ of the solution in the tube 
bringing species $k$ into the reactor and the exit flux $\phi_{\rm out} = \int_{\partial V_{\rm out}} {\bf v}\cdot d{\bf A}$ of the stirred solution.  
These fluxes are in units of m$^3$ per second, and depend on the section areas of the injection and exit tubes. 
The volume of the solution inside the reactor being preserved, we have that $\phi_{\rm out} = \sum_k \phi_{k,{\rm in}}$.
Since the concentrations are uniform inside the reactor, Eq.~(\ref{int-eq-conc}) divided by the volume $V$ becomes
\be
\frac{dc_k}{dt} = \sum_i \nu_{ki} \, w_i + \frac{1}{\tau} (c_{k0}-c_k) \, ,
\label{eq-CSTR}
\ee
where $\tau\equiv V/\phi_{\rm out}$ is the mean residence time of the species inside the reactor, and
\be
c_{k0}\equiv \frac{\phi_{k,{\rm in}}}{\phi_{\rm out}}\, c_{k,{\rm in}}
\ee
are the injected concentrations of reactants reported to the whole volume.  
Both the residence time $\tau$ and the injected concentrations $c_{k0}$ are control parameters.  

The evolution equations for the concentrations form a set of ordinary differential equations, which are typically nonlinear.
In the limit where the residence time becomes very long, the last term of Eq.~(\ref{eq-CSTR}) becomes negligible and we recover the kinetic equations in a closed reactor, the so-called batch reactor,\cite{EP98} in which case the concentrations will sooner or later reach their equilibrium value.  In the other limit where the residence time is very short, the last term dominates so that the concentrations remain nearly equal to their value at injection: $c_k\simeq c_{k0}$.  In between, the concentrations may manifest a rich variety of different stationary, oscillatory, or chaotic behaviors in some autocatalytic or cross-catalytic reaction networks.\cite{BPV84,VL88,S91,N95,EP98}

\subsection{Thermodynamics}

A typical CSTR is functioning under atmospheric pressure and at room temperature if the reactions are not too exothermic.  
Under these conditions, the relevant thermodynamic potential is Gibbs' free energy $G$.  We assume local thermodynamic equilibrium 
for every element of the solution and consider the free energy density:
\be
g_V =  \sum_k \mu_k \, c_k \,  ,
\label{gV}
\ee
where $\mu_k$ is the chemical potential of species $k$.  

Using Eq.~(\ref{gV}) together with Gibbs' fundamental relation per unit volume
\be
d g_V= -s_V dT + dP + \sum_k \mu_k dc_k  \, ,
\ee 
where $s_V$ is the entropy density, $T$ the temperature, and $P$ the pressure, one obtains the Gibbs-Duhem relation
\be
s_V dT - dP + \sum_k c_k d\mu_k = 0 \, .
\label{GD}
\ee
Using Eq.~(\ref{GD}) under isothermal and isobaric conditions, one finds that
\be
\sum_k c_k d\mu_k=0.
\label{GD1}
\ee

Since the solution is well stirred, it is quasi homogeneous in the bulk of the tank, and 
the time evolution of the Gibbs free energy follows that of the concentrations of the various species.
Using Eqs.~(\ref{gV})-(\ref{GD1}), one obtains
\be
\frac{d g_V}{dt} = \sum_k \mu_k \frac{dc_k}{dt} ,
\ee

Now, using Eq.~(\ref{eq-CSTR}) for the concentrations,  
the time evolution of the free energy density then becomes
\be
\frac{dg_V}{dt} = \sum_{ki} \mu_k \nu_{ki} w_i + \frac{1}{\tau} \sum_k \mu_k \left( c_{k0} - c_k \right) \, .
\label{dgV/dt}
\ee

According to the mass action law,\cite{N95} the reaction rates are proportional to the concentrations of all the species entering in the reaction.  
It is convenient to make the distinction between the forward and reversed reactions so that
\be
w_{\pm i} = k_{\pm i} \prod_k \left(\frac{c_k}{c^0}\right)^{\nu_{k i}^{(\pm)}} \, ,
\ee
where $k_{\pm i}$ are the rate constants, $\nu_{ki}^{(\pm)}$
the numbers of molecules entering the forward or the reversed reaction, and $c^0$ the standard concentration of one mole per liter. The stoichiometric coefficient is thus given by $\nu_{ki}=\nu_{ki}^{(-)}-\nu_{ki}^{(+)}$, while $w_i=w_{+i}-w_{-i}$.  In a dilute solution, 
the chemical potentials of the solute species are given by $\mu_k=\mu_k^0+RT\ln(c_k/c^0)$ where $R$ is the molar gas constant.  Now, the ratio of the rate constants is related to the standard free energy of the reaction according to
\be
\frac{k_{+i}}{k_{-i}} = \exp\Big(-\sum_k \frac{\mu_k^0\nu_{ki}}{RT}\Big) \, .
\ee
The entropy production rate of the reactions is given by
\ba
\sigma &=& -\frac{1}{T} \sum_{ki} \mu_k \nu_{ki} w_i, \nonumber\\
       &=& R \sum_i \left( w_{+i}- w_{-i}\right) \ln\frac{w_{+i}}{w_{-i}} \geq 0 \, ,
\label{EPR}
\ea
which is always non-negative.

Now, combining Eq.~(\ref{EPR}) with Eq.~(\ref{dgV/dt}), the time evolution of the free energy density becomes
\be
\frac{dg_V}{dt} =-T \sigma + \frac{1}{\tau} \left( \gamma_0 - g_V \right) \, ,
\label{dgV/dt-bis}
\ee
where we have introduced the following quantity 
\be
\gamma_0=\sum_k \mu_k \, c_{k0}.
\ee
In a closed reactor where $\tau$ is infinite, the free energy will decrease towards its minimal value.  
However, in an open reactor where $\tau$ is finite, the free energy does not need to reach its minimal value. 
In this regard, nonequilibrium stationary, oscillatory, or chaotic regimes can be sustained in an open reactor.\cite{P67,N95}

The term $\left( \gamma_0 - g_V \right)/\tau$ in Eq.~(\ref{dgV/dt-bis}) has no definite sign, except in a stationary state
where it is equal to the dissipation produced by the chemical reactions and therefore must be positive. In this case, 
it is sufficient to know the Gibbs free energies of incoming and outgoing chemical species in order to know the dissipation associated with chemical reactions within the reactor. 

\subsection{General properties of reaction networks in a CSTR}
\label{RN-CSTR}

Reaction network theory allows us to obtain key properties such as the conservation laws and the cycles, which determine the behavior of the stationary states. These properties are known for chemostatted systems,\cite{RE16,Polettini2014} and an important issue is to understand how they differ in a CSTR. We note that the cycles defined in reaction network theory should not be confused with the limit cycles of nonlinear dynamics. The former are defined as the right null eigenvectors of the stoichiometric matrix,\cite{RE16} while the latter are periodic solutions for the ordinary differential equations of the reaction network corresponding to periodic oscillations.\cite{NP77,N95}

The equations (\ref{eq-CSTR}) ruling the time evolution of the concentrations can be rewritten in matrix form as follows:
\be
\frac{d{\bf c}}{dt} = \pmb{\nu}\cdot{\bf w} + \frac{1}{\tau}\left( {\bf c}_0-{\bf c}\right) ,
\label{eq-CSTR-v}
\ee
in terms of the $s$-dimensional vectors $\bf c$ and ${\bf c}_0$ of concentrations and injected concentrations, the $r\times s$ matrix $\pmb{\nu}$ of stoichiometric coefficients, and the $r$-dimensional vector of reaction rates $\bf w$, where $s$ is the number of species and $r$ the number of reactions in the network.

In the limit $\tau \to \infty$, we recover the case of a closed reactor.\cite{RE16,Polettini2014}
In a stationary state, we have 
$\pmb{\nu}\cdot{\bf w} =0,$
implying that ${\bf w}$ can be decomposed in the basis of right null eigenvectors ${\bf e}_\gamma$, which are called cycles:
${\bf w}= \sum_\gamma w_\gamma {\bf e}_\gamma$.

The rank of the stoichiometry matrix of the closed reactor can be written as 
\be
{\rm rank}(\pmb{\nu}) = r - o= s - l \, ,
\label{rank-closed}
\ee
where $o={\rm dim \, ker}(\pmb{\nu})$ is the number of cycles, and $l={\rm dim\, coker}(\pmb{\nu})$ 
the number of conserved quantities.
The quantities that are conserved in a closed reactor are defined as
\be
L \equiv \pmb{\ell}\cdot{\bf c} \, ,
\ee
with a vector $\pmb{\ell}$ such that
\be
\pmb{\ell}\cdot\pmb{\nu} = 0 \, .
\ee
In an open reactor where $\tau$ is finite, such quantities are no longer conserved. 
Instead, they converge asymptotically towards their value defined for the injected concentrations:
\be
\label{L0}
L_0 = \pmb{\ell}\cdot{\bf c}_0 \, .
\ee
Indeed, applying the vector $\pmb{\ell}$ to Eq.~(\ref{eq-CSTR-v}), we find that
\be
\label{L-relaxation}
\frac{dL}{dt} = \frac{1}{\tau} (L_0-L) \, ,
\ee
the solution of which is given by
\be
L(t) = L(0) \, {\rm e}^{-t/\tau} + L_0 \left( 1 - {\rm e}^{-t/\tau}\right) .
\ee
It is important to emphasize that all conservation laws are broken in a CSTR. 

We can also recover this result using the full 
stoichiometry matrix of the CSTR.
In an open reactor, the reaction network also includes the $s$ reactions of rates $\left( {\bf c}_0-{\bf c}\right)/\tau$ 
so that the total number of reactions becomes $r'=r+s$, while the matrix of stoichiometric coefficients should be extended 
towards a $r'\times s$ matrix with $r'=r+s$.
This means that the new stoichiometry matrix of the CSTR reads 
\be 
\pmb{\nu}'= \left( \pmb{\nu}, \pmb{I} \right), 
\label{def-nu'}
\ee
where 
$\pmb{I}$ is the identity matrix $s \times s$. Therefore Eq.~(\ref{eq-CSTR-v}) becomes
\be
\frac{d{\bf c}}{dt} = \pmb{\nu}'\cdot{\bf w'} ,
\label{eq-CSTR-ext}
\ee
with the flow rate ${\bf w'}=\left( {\bf w}, {\bf \tilde{w}} \right)^{\rm T}$ a column matrix of dimension $1\times r'$ with ${\bf \tilde{w}}=( {\bf c}_0-{\bf c} )/\tau$.

In an open reactor, we also get
\be
{\rm rank}(\pmb{\nu}') = r' - {\rm dim\, ker} (\pmb{\nu}')= s - {\rm dim\, coker} (\pmb{\nu}') \, .
\label{rank-open}
\ee
The number of conserved quantities is now equal to zero $l'={\rm dim\, coker} (\pmb{\nu}')=0$ so that the number of cycles is equal to the
 number of reactions in the original network: $o'={\rm dim\, ker} (\pmb{\nu}')=r$. 
Therefore, there are $o'-o=r-o=s-l$ cycles of the open network
that were not already present in the corresponding closed network. For chemostatted systems, such cycles have been called emergent cycles.\cite{Polettini2014,RE16}

Here, we choose to call these cycles external cycles, because they involve the flow rates ${\bf \tilde{w}}$ which are specific to the CSTR.  The other cycles are called internal. A general cycle $\bf{e}'$ can be split into network components and flow components as $\bf{e}'=(\bf{e},{\bf \tilde{e}})^{\rm T}$. This cycle obeys $\pmb{\nu}' \cdot {\bf e}'=\pmb{\nu} \cdot {\bf e} +  {\bf \tilde{e}}=0$. Here we can make the distinction between internal cycles ${\bf e}_\gamma$ previously defined for the network of the closed reactor which are such that
$\pmb{\nu} \cdot {\bf e}_\gamma=0$ and ${\bf \tilde{e}}_\gamma=0$; and external cycles ${\bf e}_\alpha$ which are such that ${\bf \tilde{e}}_\alpha=-
\pmb{\nu} \cdot {\bf e}_\alpha \neq 0$. 

As far as the thermodynamic description of the system is concerned, 
Eq.~(\ref{dgV/dt-bis}) becomes 
\be
\frac{dg_V}{dt} = \sum_{ki} \mu_k \nu'_{ki} w'_i \, ,
\ee
within the framework of the extended network.
In a stationary state, the entropy production rate of Eq.~(\ref{EPR}) may be rewritten as:
\ba
\sigma &=& -\frac{1}{T} \, \pmb{\mu} \cdot \pmb {\nu} \cdot {\bf w} = -\frac{1}{T} \sum_\lambda w_\lambda \, \pmb{\mu} \cdot \pmb{\nu} \cdot {\bf e}_\lambda, \nonumber\\
          &=& -\frac{1}{T} \sum_\alpha w_\alpha \, \pmb{\mu} \cdot \pmb {\nu} \cdot {\bf e}_\alpha, \nonumber\\
	  &=& \frac{1}{T} \sum_\alpha w_\alpha \, \pmb{\mu} \cdot {\bf \tilde{e}}_\alpha \geq 0.
\ea 
This shows that in this case the entropy production rate can be written as a sum of contribution from external cycles denoted with the index $\alpha$ only. A similar property was reported in the case of chemostatted systems.\cite{Polettini2014,RE16}

\subsection{Illustrative examples}

Here, we present two illustrative examples of the above framework. 
The first example is a network of small size taken from 
Ref.~\onlinecite{Polettini2014}, and the second one is a larger network 
describing polymers with a mass-exchange process taken from Ref.~\onlinecite{Rao2015a}.

\subsubsection{Example with a finite network}

\begin{figure}[h!]
\includegraphics[scale=0.6]{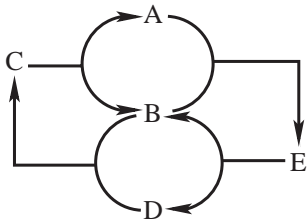}
\caption{Hypergraph of the closed chemical network (\ref{w1})-(\ref{w4}).}
\label{fig1.5}
\end{figure}

The set of reactions in the first example are
\ba
&&\ce{A + B <=>[1] C},  \qquad \qquad w_1 = k_{+1} [\ce{A}] [\ce{B}] - k_{-1} [\ce{C}], \label{w1}\\ 
&&\ce{C <=>[2] B + D}, \qquad \qquad w_2 = k_{+2} [\ce{C}] - k_{-2} [\ce{B}] [\ce{D}], \label{w2}\\ 
&&\ce{B + D <=>[3] E}, \qquad \qquad w_3 = k_{+3} [\ce{B}] [\ce{D}] - k_{-3} [\ce{E}], \label{w3}\\
&&\ce{E <=>[4] A + B}, \qquad \qquad w_4 = k_{+4} [\ce{E}] - k_{-4} [\ce{A}] [\ce{B}]. \label{w4}
\ea
The stoichiometry matrix of this network is then
\ba
\pmb{\nu}=
\left(
\begin{array}{rrrr}
 -1 & 0 & 0 & 1 \\
-1 & 1 & -1 & 1 \\
1 & -1 & 0 & 0 \\
0 & 1 & -1 & 0 \\
0 & 0 & 1 & -1 
\end{array}
\right), 
\ea
and the corresponding hypergraph is shown in Fig.~\ref{fig1.5}.\cite{RE16}

As shown in Ref.~\onlinecite{Polettini2014}, this network has $l=2$ conserved 
quantities $L_1= [\ce{B}] + [\ce{C}] + [\ce{E}]$ 
and $L_2 = [\ce{A}] + [\ce{C}] + [\ce{D}] + [\ce{E}]$.  There is only one cycle ($o=1$), with a null right 
eigenvector $(1,1,1,1)^{\rm T}$.

\begin{figure}[h!]
\includegraphics[scale=0.4]{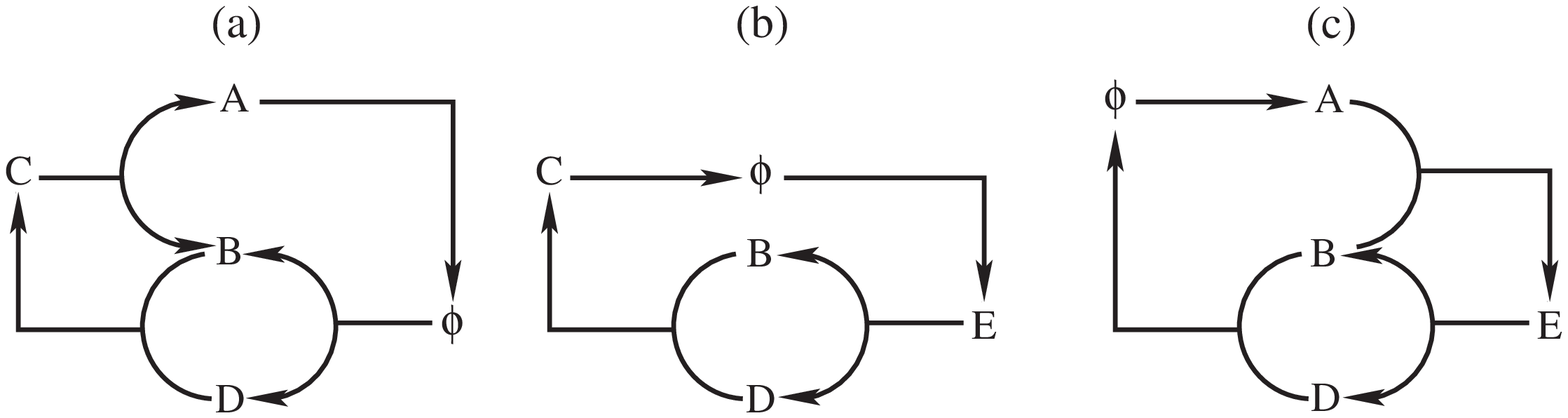}
\caption{Hypergraphs of the three new cycles in the open version of the chemical network represented in Fig.~\ref{fig1.5}. 
Here $(a)$, $(b)$ and $(c)$ correspond to the cycles ${\bf e}_2$, ${\bf e}_3$ and ${\bf e}_4$ respectively. 
Note the appearance of the symbol $\phi$ which is a notation for new reactions involving the inflow and outflow of the CSTR.}
\label{fig1.7}
\end{figure}

For the open reactor network, the stoichiometry matrix $\pmb{\nu}'$ is obtained from 
Eq.~(\ref{def-nu'}), its rank is 5, it has $o'=4$ cycles, 
$l'=0$ conserved quantities, and $s-l=3$ new cycles. 
The cycles are the old cycle  ${\bf e}_1=(1,1,1,1,0,0,0,0,0)^{\rm T}$ and the three new cycles 
${\bf e}_2=(1,1,0,0,1,0,0,-1,0)^{\rm T}$, ${\bf e}_3=(0,1,1,0,0,0,1,0,-1)^{\rm T}$, 
and ${\bf e}_4=(0,0,1,1,-1,0,0,1,0)^{\rm T}$. The new cycles are represented in Fig.~\ref{fig1.7}.
This representation makes it clear that hypergraphs depicting the new cycles of the open network are built from the hypergraph of the closed network by removing some reactions and chemical species. Then the remaining pieces are connected together using a special symbol $\phi$, which is introduced for this purpose and which describes new reaction pathways involving the exterior of the CSTR.

We note that the hypergraphs in Figs. \ref{fig1.5} and \ref{fig1.7} depend on the reaction network, but not on the concentration values of the involved species.

\subsubsection{Example with an infinite network}

We now move to a more complex reaction network, namely the 
model of polymers undergoing a mass-exchange process taken from Ref.~\onlinecite{Rao2015a}.
In this model, two polymers of mass $n$ and $m$ interact with the 
reaction

\be
(n) + (m) \ce{<=>[\kappa]} (n + 1) + (m - 1),  \qquad {\rm for} \quad n \ge 1, \ m \ge 2.
\ee

In an open reactor, the kinetic equations can be written in the form:
\be
\frac{dc_k}{dt} = \frac{1}{2} \sum_{n\geq 1, m\geq 2} \nu_{k,nm} \, w_{nm} + \frac{1}{\tau} (c_{k,0}-c_k) \quad \quad {\rm for} \quad k \ge 1, 
\label{mass-exchange_kin_eqs}
\ee
with the stoichiometric coefficients $\nu_{k,nm}=\delta_{k,n+1}+\delta_{k,m-1}-\delta_{k,n}-\delta_{k,m}$ and the rates $w_{nm}=\kappa c_nc_m-\kappa c_{n+1}c_{m-1}$ obeying the mass action law.  

In the closed reactor ($\tau=\infty$), this network has two conserved quantities: the total concentration $c\equiv \sum_{k=1}^{\infty} c_k$ and the total number of monomeric units $M=\sum_{k=1}^{\infty} k \, c_k$.  In the open reactor, these quantities are no longer conserved because they obey the equations
\bea
\frac{dc}{dt} &=& \frac{1}{\tau}\left( c_0-c\right) \, , \\
\frac{dM}{dt} &=& \frac{1}{\tau}\left( M_0-M\right) \, , 
\label{Evolution of cons-laws}
\eea
so that they converge asymptotically in time towards their value $c_0$ or $M_0$ fixed by the inlet concentrations.

Although the reaction network is infinite, it can be truncated by considering a finite number $s$ of species.  In this case, the reactions and the cycles can be enumerated using the list of all the reactions:
\bea
&&\ce{1 + 2 <=> 2 + 1} \, , \quad  \ce{2 + 3 <=> 3 + 2} \, , \quad  \ce{3 + 4 <=> 4 + 3} \, , \ \dots \nonumber\\
&&\ce{1 + 3 <=> 2 + 2} \, , \quad  \ce{2 + 4 <=> 3 + 3} \, , \ \dots \nonumber\\
&&\ce{1 + 4 <=> 2 + 3} \, , \ \dots \nonumber\\
&& \qquad\quad\vdots
\eea
In the closed reactor, the number of reactions involving $s$ species is thus equal to
\be
r=\frac{1}{2}  s(s-1) \, .
\ee
Since there are $l=2$ conserved quantities in the closed reactor, Eq.~(\ref{rank-closed}) 
thus shows that the number of cycles is equal to
\be
o=r-s+2=\frac{1}{2}  (s-1)(s-2)+1 \, .
\ee
Accordingly, these numbers are increasing quadratically with the number $s$ of species.

In the open reactor, the reactions include the rates $\tilde w_k=(c_{k0}-c_k)/\tau$ due to the flow so that the number of reactions involving $s$ species is now given by
\be
r' = r+s = \frac{1}{2}  s(s+1) \, .
\ee
There are no conserved quantities $l'=0$ and the number of cycles is here equal to
\be
o' = o+s-2 = r = \frac{1}{2}  s(s-1) \, .
\ee
Therefore, opening the reactor only adds a number of new cycles $s-2$ that is increasing 
linearly with the number of species, while the total number of cycles of the open system is increasing
quadratically with the number of species.

\begin{figure}[h!]
\includegraphics[scale=0.5]{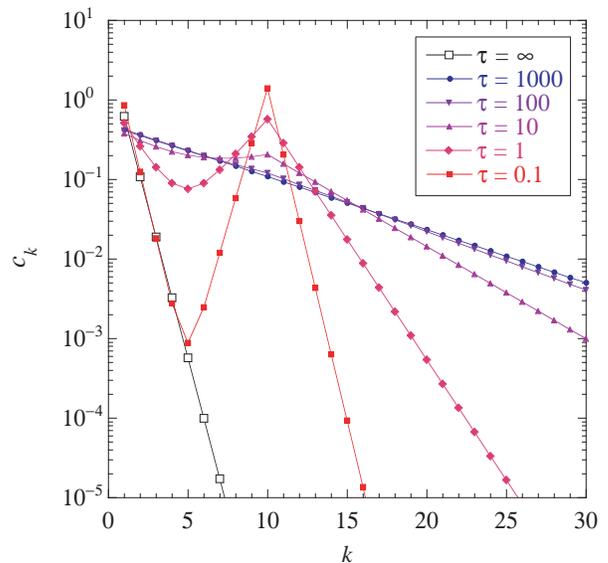}
\caption{Stationary distributions of the oligomer concentrations $\{c_k\}$ for the mass-exchange process
 with the rate constant $\kappa=1$ in a CSTR with the injection of monomers and 10-mers at the inlet 
concentrations $c_{1,0}=1$ and $c_{10,0}=2$ for different 
values of the residence time $\tau$.  If $\tau=\infty$, the reactor is closed and the stationary
distribution is the equilibrium one (open squares).  If $\tau$ is finite, the reactor is open and out 
of equilibrium (filled symbols).}
\label{fig2}
\end{figure}

In the CSTR, all the concentrations remain bounded in time. 
This rules out the possibility to observe an ``unbalanced phase'', such as the unbounded growth 
phase reported in Ref.~\onlinecite{Rao2015a} in a variant of this mass-exchange model, 
which was driven out-of-equilibrium by chemostats fixing the concentrations of polymers of certain lengths. In that model, the total concentration 
$c$ increased linearly in time and the total number of monomers $M$ increased quadratically. In contrast, in a CSTR both quantities remain bounded in time, a property which follows generally from Eq.~(\ref{L-relaxation}).

In Fig.~\ref{fig2}, we show the stationary distribution of concentrations in a CSTR for different 
values of the residence time $\tau$ by injecting monomers at the concentration $c_{1,0}$ and 
oligomers of length $l=10$ at the concentration $c_{10,0}$.  The kinetic equations are integrated with a Runge-Kutta algorithm of orders 4 and 5 with variable steps from the initial distribution $c_k(0)=\exp(-k^2/2)$.  The distribution is plotted after a time interval $t=1000$ if $\tau=\infty,0.1, 1, 10$, after $t=10000$ if $\tau=100$, and after $t=50000$ if $\tau=1000$, when stationarity is numerically
 reached.  If $\tau=\infty$, the reactor is closed so that the concentrations reach their equilibrium exponential distribution
\be
c_{k,{\rm eq}} = \frac{c(0)^2}{M(0)} \left[ 1 - \frac{c(0)}{M(0)}\right]^{k-1} ,
\label{ME_equil}
\ee
determined by the initial values of the two invariant quantities $c(0)=0.7533$ and $M(0)=0.9119$, so that $c_{k,{\rm eq}} =3.58\times 0.174^k$. 
 In contrast, under nonequilibrium conditions if $\tau$ is finite, the distribution deviates from being purely exponential and it even becomes
 bimodal with peaks at $k=1$ and $k=10$ if the open reactor is strongly out of equilibrium with a small enough residence time~$\tau$.  
Nevertheless, the distribution is always exponential beyond the largest injected concentration $c_{10,0}$, as shown in Appendix~\ref{AppA}. 
 In the open reactor, the distribution no longer depends on the initial conditions but on the values of the injected 
concentrations.

\section{Serial transfers between closed reactors}
\label{sec:ST}

Now, we consider the dynamics of the reaction network in a typical serial transfer experiment.\cite{Vaidya2012}

\subsection{Time evolution of the concentrations}

\begin{figure}[h!]
\includegraphics[scale=0.4]{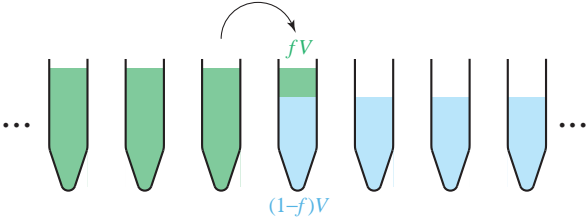}
\caption{Schematic representation of a serial transfer experiment in which a volume $fV$ of the solution of interest (green) is transferred repeatedly 
into a fresh solutions of volume $(1-f)V$ (blue).}
\label{fig5}
\end{figure}

At every transfer, a fraction $f$ of the solution volume $V$ is transferred to another 
closed reactor already containing a fresh solution of volume $(1-f)V$ with reactants 
at the concentrations $c_{k0}$ as illustrated in Fig.~\ref{fig5}. 

Let $\mathscr{T}$ be the time interval between two transfers.  
During this time interval, the reactor is closed so that
 the concentrations evolves according to
\be
\frac{d{\bf c}}{dt} = \pmb{\nu}\cdot{\bf w} \, .
\label{eq-closed}
\ee
Let ${\bf c}(n{\mathscr T}-0)$ be the concentrations just before the previous transfer.
The concentrations just after the transfer and stirring are thus given by
\be
{\bf c}(n{\mathscr T}+0)= (1-f) \, {\bf c}_0 + f \, {\bf c}(n{\mathscr T}-0) \, .
\ee
Thereafter, the concentrations evolves according to
\be
{\bf c}\left( t\right) = {\bf c}(n{\mathscr T}+0) + \int_{n{\mathscr T}}^{t} \pmb{\nu}\cdot{\bf w}\left[{\bf c}(t')\right] \, dt' 
\ee
with $n{\mathscr T}+0 <t< n{\mathscr T}+{\mathscr T}-0$.
The concentrations just before the next transfer are thus given by
\bea
{\bf c}(n{\mathscr T}+{\mathscr T}-0) &=& (1-f) \, {\bf c}_0 + f \, {\bf c}(n{\mathscr T}-0) \nonumber\\
&&+ \int_{n{\mathscr T}}^{(n+1){\mathscr T}} \pmb{\nu}\cdot{\bf w}\left[{\bf c}(t)\right] \, dt \, ,
\label{map}
\eea
which defines a mapping ${\bf c}_{n+1}=\pmb{\Phi}({\bf c}_n)$ from ${\bf c}_n\equiv {\bf c}(n{\mathscr T}-0)$ to ${\bf c}_{n+1}\equiv{\bf c}(n{\mathscr T}+{\mathscr T}-0)$. A similar mapping can be obtained for the concentrations after the transfers.

Let us suppose that the transfers are quickly repeated every small time interval ${\mathscr T}=\Delta t$.  As a consequence of Eq.~(\ref{map}), we get the approximate ordinary differential equations:
\be
\frac{\Delta{\bf c}}{\Delta t} \simeq \frac{1-f}{\Delta t} \, ({\bf c}_0 -{\bf c}_n) +  \pmb{\nu}\cdot{\bf w}\left[(1-f) \, {\bf c}_0 + f \, {\bf c}_n\right] \, ,
\ee
where $\Delta{\bf c}={\bf c}_{n+1}-{\bf c}_n$. Introducing the effective residence time 
\be
\label{residence_time}
\tau \equiv \frac{\Delta t}{1-f} \, ,
\ee
we recover in the limit $\Delta t\to 0$ the kinetic equations of the concentrations in a CSTR:
\be
\frac{d{\bf c}}{dt} = \pmb{\nu}\cdot{\bf w}({\bf c}) + \frac{1}{\tau} \left( {\bf c}_0 - {\bf c}\right) \, .
\label{eq-open}
\ee
If $f=1-{\mathscr T}/\tau$ in the limit ${\mathscr T}\to 0$, an experiment of serial transfers between closed reactors is thus similar to an experiment in a CSTR. Therefore, similar nonequilibrium regimes are expected in both experiments under comparable conditions.

\subsection{Thermodynamics}

Let us follow Gibbs' free energy during the time evolution.  Before the transfer at time $n{\mathscr T}$, the free energy 
density of the solution in the volume $V$ is $g_V\left[{\bf c}\left(n{\mathscr T}-0\right) \right]$.
After the transfer of the volume $fV$ of solution into the volume $(1-f)V$ of fresh solution and the mixing of both, the free energy density becomes
$g_V\left[ f {\bf c}\left(n{\mathscr T}-0\right) + (1-f){\bf c}_0\right]$.
Thereafter, the free energy density changes in time since the concentrations evolve according to Eq.~(\ref{eq-closed}) in the closed reactor.  
At the end of the time interval $n{\mathscr T}<t<n{\mathscr T}+{\mathscr T}$, the free energy density has thus become
\bea
 g_V\left[{\bf c}\left(n{\mathscr T}+{\mathscr T}-0\right) \right] &=& g_V\left[ f {\bf c}\left(n{\mathscr T}-0\right) + (1-f){\bf c}_0\right] \nonumber\\
 &&+ \int_{n{\mathscr T}}^{(n+1){\mathscr T}} dt \, \dot{g}_V\left[{\bf c}(t)\right],
\label{G+1}
\eea
where $\dot{g}_V=\pmb{\mu}\cdot\pmb{\nu}\cdot{\bf w}$ is the time derivative of the free energy in the closed reactor given by Eq.~(\ref{dgV/dt}) with $\tau=\infty$.
The process repeats itself at every time interval.

In the limit where ${\mathscr T}=\Delta t\to 0$ with $f=1-\Delta t/\tau$, using the same notation ${\bf c}_n\equiv {\bf c}(n{\mathscr T}-0)$ as above, Eq.~(\ref{G+1}) becomes
\bea
g_V({\bf c}_{n+1}) &=& g_V\left[ {\bf c}_n + \frac{\Delta t}{\tau}({\bf c}_0-{\bf c}_n)\right] \nonumber\\
&& + \Delta t \, \pmb{\mu}({\bf c}_n)\cdot\pmb{\nu}\cdot{\bf w}({\bf c}_n)+O(\Delta t^2)\, .
\eea
Since $\pmb{\mu}=\partial g_V/\partial{\bf c}$, the previous equation becomes
\bea
g_V({\bf c}_{n+1}) &=& g_V({\bf c}_n) + \frac{\Delta t}{\tau}\, \pmb{\mu}({\bf c}_n)\cdot ({\bf c}_0-{\bf c}_n) \nonumber\\
&& + \Delta t \, \pmb{\mu}({\bf c}_n)\cdot\pmb{\nu}\cdot{\bf w}({\bf c}_n)+O(\Delta t^2)\, .
\eea
In the limit $\Delta t\to 0$, we thus find the differential equation
\be
\frac{dg_V}{dt} = \pmb{\mu}({\bf c})\cdot\pmb{\nu}\cdot{\bf w}({\bf c}) + \frac{1}{\tau}\, \pmb{\mu}({\bf c})\cdot ({\bf c}_0-{\bf c}) \, ,
\ee
which is the same as Eq.~(\ref{dgV/dt}) for the time evolution of the free energy in the CSTR.  

In the limit $\Delta t\to 0$, there is thus equivalence between the dynamics in the CSTR and the time evolution in a serial transfer experiment.

\subsection{General properties of the reaction network in serial transfers}

The considerations of Subsec.~\ref{RN-CSTR} extends to reaction networks in serial transfers between closed reactors. Here,  a stationary state corresponds to a fixed point ${\bf c}_n={\bf c}_*=\pmb{\Phi}({\bf c}_*)$ of the mapping defined by Eq.~(\ref{map}). 

As in the case of the CSTR, conserved quantities of the closed network, namely quantities of the form $L=\pmb{\ell}\cdot{\bf c}$ are no longer conserved in the open reactor. Instead, their dynamics follows a simple relaxation equation
\be
L\left(n{\mathscr T}+ {\mathscr T} -0 \right) = (1-f) L_0 + f L \left( n {\mathscr T} -0 \right),
\ee
which is the counterpart of Eq.~(\ref{L-relaxation}). At the fixed point where the conserved quantity is such that $L\left( n {\mathscr T}+ {\mathscr T} +0 \right)=L \left( n {\mathscr T} +0 \right)=L_*$, this quantity equals the quantity $L_0$, which is the conserved quantity of the closed network evaluated at the injected concentration and which was introduced in Eq.~(\ref{L0}).

Furthermore, the fixed point ${\bf c}_*$ should satisfy the same condition
\be
\pmb{\nu}'\cdot{\bf w}'=0 \, , 
\ee 
as in Subsec.~\ref{RN-CSTR} in terms of the same stoichiometry matrix~(\ref{def-nu'}), which was introduced to characterize the CSTR. Note however that now ${\bf w}'$ is replaced by ${\bf w}'=(\langle{\bf w}\rangle,\tilde{\bf w})^{\rm T}$ with the time average of the reaction rates over the time interval between the transfers
\be
\langle{\bf w}\rangle = \frac{1}{{\mathscr T}} \int_{n{\mathscr T}}^{(n+1){\mathscr T}} {\bf w}\left[{\bf c}(t)\right] \, dt \, ,
\ee
which has the same value between every transfer because the process repeats itself from the point fixed ${\bf c}_n={\bf c}_*$, and
\be
\tilde{\bf w} = \frac{1-f}{{\mathscr T}} \left( {\bf c}_0-{\bf c}_*\right) .
\ee
Therefore, Eq.~(\ref{rank-open}) applies here as well and the number of conserved quantities is equal to zero.  The rates can be decomposed as ${\bf w}'=\sum_\lambda w'_\lambda {\bf e}'_\lambda$ onto the $o'={\rm dim\, ker}(\pmb{\nu}')$ right null eigenvectors of the matrix $\pmb{\nu}'$, which define the cycles, as in Subsec.~\ref{RN-CSTR}.

\subsection{Illustrative example}

\begin{figure}[h!]
\includegraphics[scale=0.5]{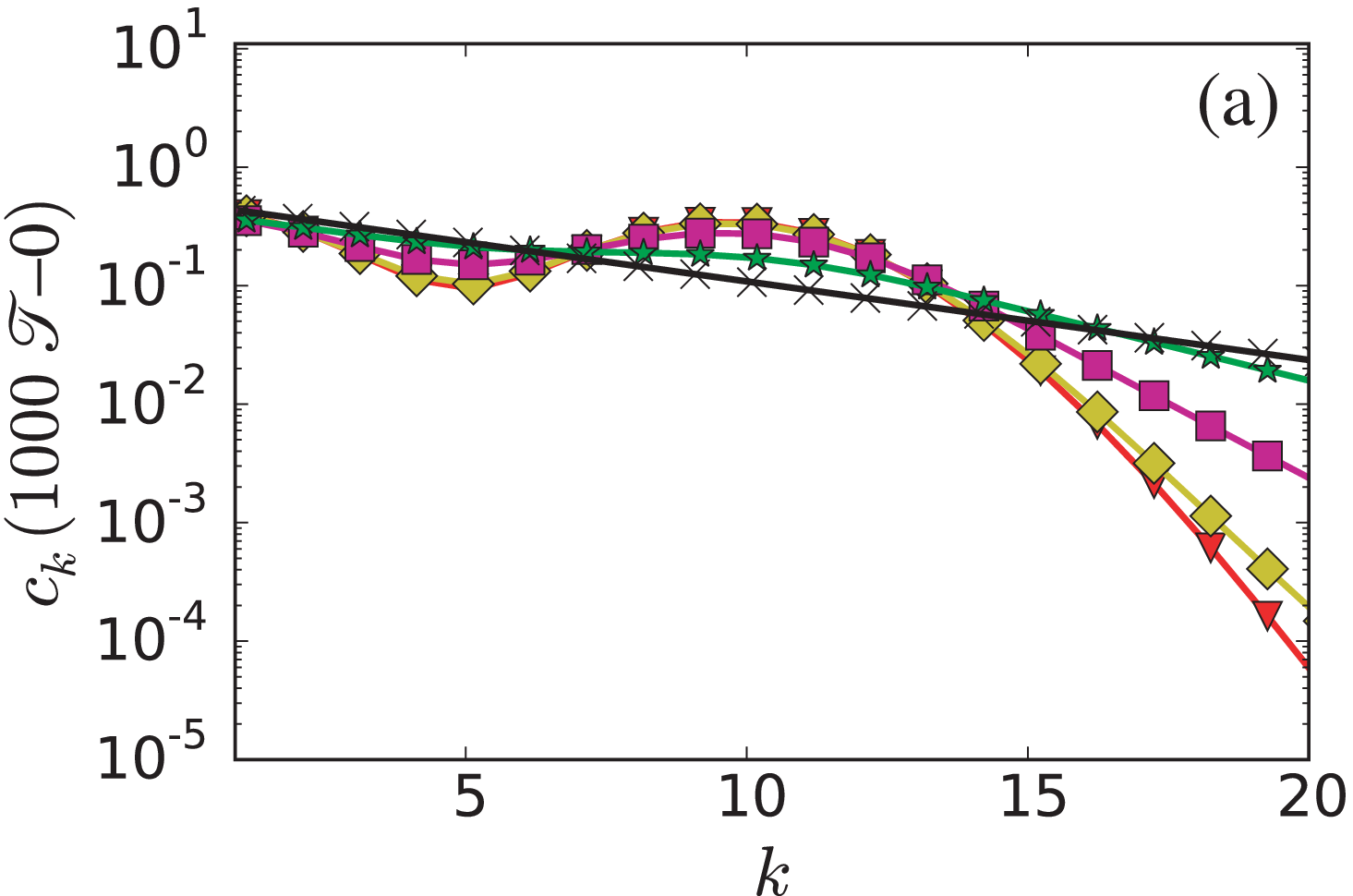}
\includegraphics[scale=0.5]{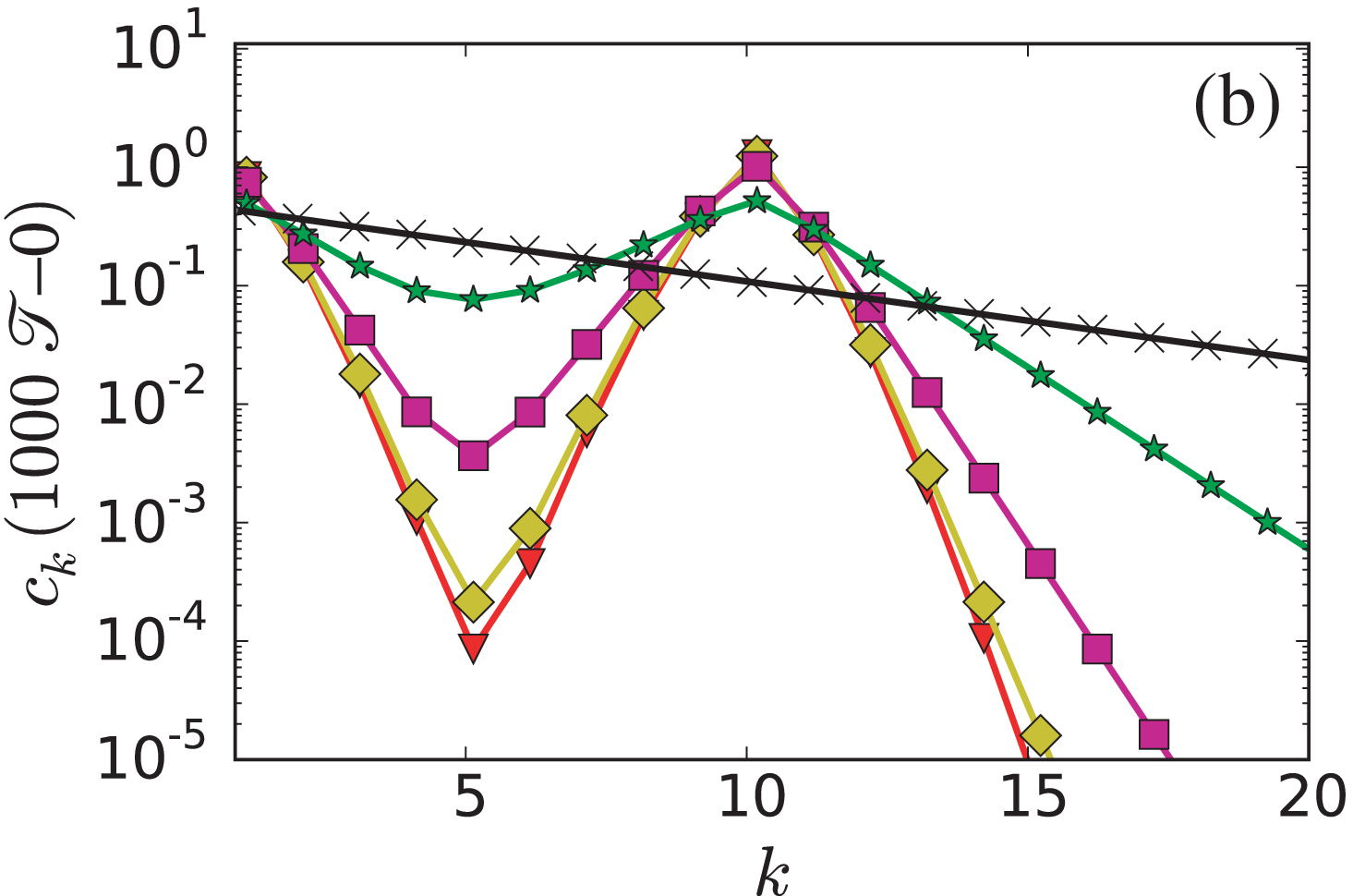}
\caption{Concentrations $c_k$ of oligomers versus their length $k$ probed at the time $1000{\mathscr T}-0$ after a thousand serial transfers with fixed parameters (a) ${\mathscr T}=1$ and (b) ${\mathscr T}=0.1$ and for various values of $f$. Symbols correspond to $f=0.01$ (downward red triangles), $f=0.1$ (yellow diamonds), $f=0.5$ (magenta squares), $f=0.9$ (green stars), and black crosses represent the equilibrium distribution.}
\label{fig3}
\end{figure}

Here, we illustrate the correspondence between the serial transfers and CSTR dynamics
 using the mass-exchange model introduced above. 
The conditions of operation of the reactors are the same as in Fig.~\ref{fig2}, namely 
monomers are injected at the concentration $c_{1,0}=1$ and 
oligomers of length $l=10$ at the concentration $c_{10,0}=2$. The main difference is that now the reactor is evolving by serial transfers instead of 
the CSTR dynamics. The kinetic equations have been integrated using the integrator {\tt odeint}, 
which is available in SciPython. The precision of this integrator is fixed to $10^{-5}$, which is 
the same as that used in Fig.~\ref{fig2}.
The length distributions of the oligomers
have been observed at the time $1000{\mathscr T}-0$, 
at which we find that the distributions have reached stationarity.
In Fig.~\ref{fig3}, simulations of serial transfers have been carried out 
keeping the time ${\mathscr T}$ fixed while varying $f$. As expected in this case, the length 
distribution approaches the equilibrium exponential distribution in the limit $f \to 1$, 
since the residence time introduced in Eq.~(\ref{residence_time}) becomes infinite.

\begin{figure}[h!]
\includegraphics[scale=0.5]{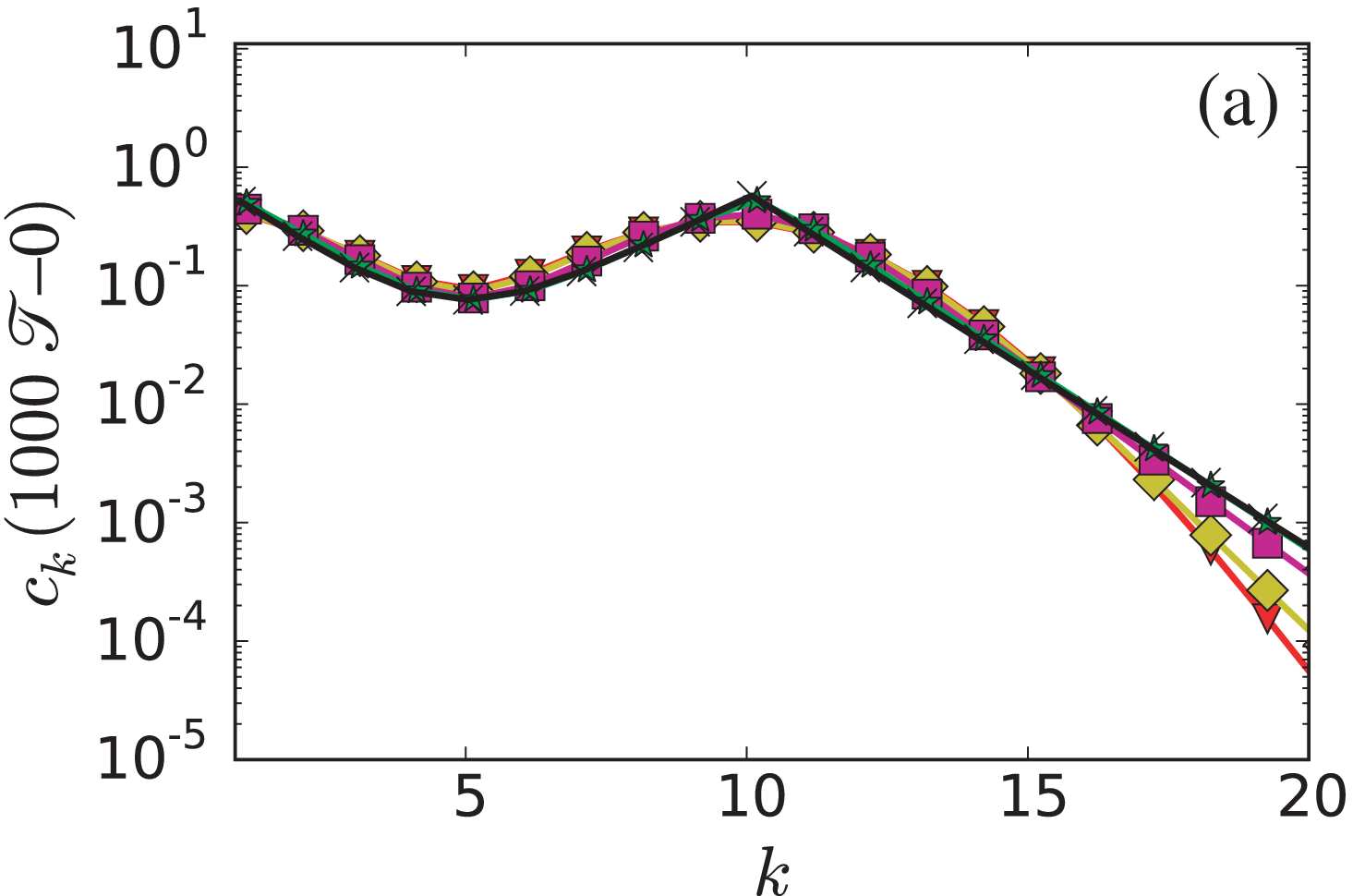}
\includegraphics[scale=0.5]{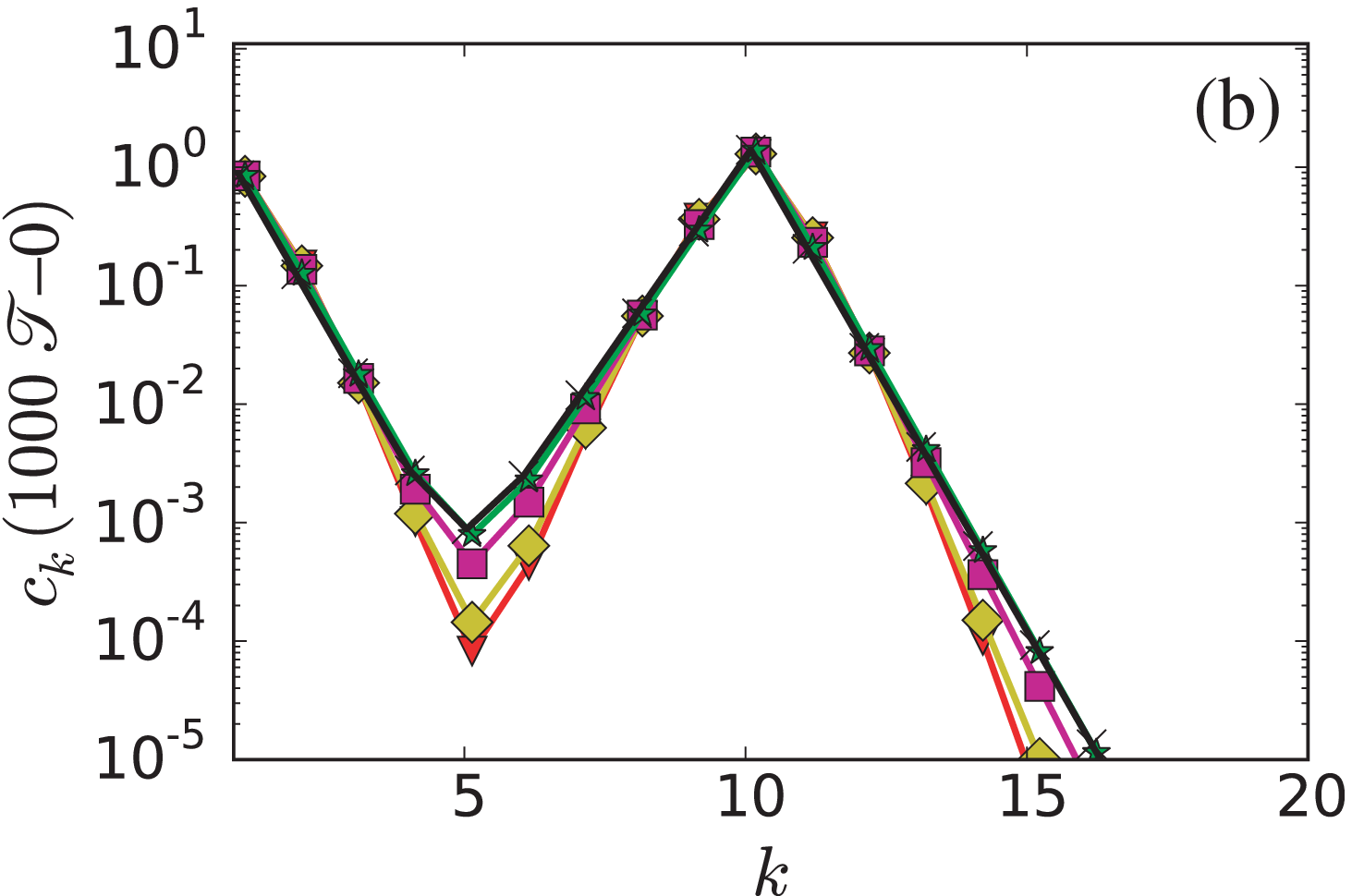}
\caption{Concentrations $c_k$ of oligomers versus their length $k$ probed at the time 
$1000{\mathscr T}-0$ after a thousand serial transfers 
corresponding to varying $({\mathscr T},f)$ parameters at fixed residence time 
(a) $\tau_{\rm eff}=1$ or (b) $\tau_{\rm eff}=0.1$.
Symbols correspond to $f=0.01$ (downward red triangles), $f=0.1$ (yellow diamonds), 
$f=0.5$ (magenta squares), $f=0.9$ (green stars), but now black crosses represent the length distribution 
predicted by the CSTR dynamics.}
\label{fig4}
\end{figure}

In order to test more precisely the convergence towards the CSTR dynamics,
we have varied in Fig.~\ref{fig4} the parameters $(f,{\mathscr T})$ 
while keeping the residence time $\tau_{\rm eff}=\tau$ constant either at the value 1 or 0.1.
The length distributions of the oligomers have been observed at the time $1000{\mathscr T}-0$.
These plots indeed confirm that, in this system, a convergence towards the CSTR is obtained 
when $f \to 1$, which is equivalent to ${\mathscr T} \to 0$ since 
the residence time $\tau$ is kept constant.

In general, the state of the reactor following serial transfers with arbitrary parameters $(f,{\mathscr T})$ can differ substantially from the predictions of the CSTR. However, if the parameters $(f,{\mathscr T})$ are chosen according to Eq.~(\ref{residence_time}) and the time of observation is not too long, as shown in Fig.~\ref{fig4}, the behavior resulting from serial transfers can be quite close to that observed by the CSTR dynamics even when the parameter $f$ is varied in a large range from 0.001 to 0.99.

\section{Conclusion}
\label{conclusion}

In this paper, we have made a comparative study of the kinetics and thermodynamics 
of open reactors (CSTR) with that of serial transfers between closed reactors.
For a given choice of a chemical network and injected species, 
both the CSTR and the serial transfer dynamics admit 
a steady state. This implies that both systems will reach comparable
composition on long times and also that the same cycles can be used to characterize the steady state of both systems. However, their dynamics can differ substantially. Only in the limit where the time interval between serial transfers tends to zero for a fixed residence time, the two dynamics are strictly equivalent.

We have also compared the properties of reaction networks in chemostatted systems\cite{Polettini2014,RE16} with those in a CSTR.  In contrast to chemostatted systems, the concentrations remain bounded in a CSTR.  In a CSTR, there is no remaining conserved quantity and new cycles involving the exterior of the CSTR appear in addition to the cycles of the the closed reactor network.  Similar results hold for the serial transfer dynamics.
 
This study was motivated by 
molecular evolution experiments which use serial transfers 
in the context of research on the origins of life.\cite{Vaidya2012}
Besides chemical evolution by serial transfers, 
experiments in this field also use a dry-wet (or day-night) cycling 
protocol as a means of inducing an evolution in the composition of the system.\cite{Segre2000,Tkachenko2015,Forsythe2017}
Many other cycling protocols are possible. In particular, cycles driven by thermal convection\cite{Mast2013} 
and cycles of activation-deactivation or of compartmentalization-decompartmentalization
of specific species\cite{Matsumura2016} are being considered.
We believe that the framework presented here could be extended to cover these cases provided 
the molecular interactions between the various species can be modeled. 

Cycling protocols are often explored in the literature in order 
to explain how a population of sufficiently long chains can be self-sustained
and show emergent properties, an important issue for the research on the origins of life.
Our example of polymerization with mass exchange indicates that 
a population of polymers with a non-exponential distribution can be maintained in an 
open reactor if the residence time is not too long, thus 
solving the first issue. The second issue regarding the emergence of new properties is 
clearly more complex, but the 
framework used here allows at least to 
identify emergent cycles of the chemical network, which should capture important features of the emergent properties we are after.


\section*{Acknowledgments}

P. Gaspard thanks the ESPCI, the Universit\'e libre de Bruxelles (ULB), and the Fonds de la Recherche Scientifique~-~FNRS under the Grant PDR~T.0094.16 of the project ``SYMSTATPHYS" for financial support. A.B. was supported by the Agence Nationale de Recherche (ANR-10-IDEX-0001-02, IRIS OCAV). D. L. would like to thank N. Lehman for stimulating discussions. 

\appendix

\section{Analysis of the mass-exchange model}
\label{AppA}

In order to determine the stationary 
concentrations of the mass-exchange model in the CSTR, we take the explicit form of the kinetic equations~(\ref{mass-exchange_kin_eqs}) with injection of monomers and oligomers of length $l$:\cite{Rao2015a}
\be
\frac{dc_k}{dt} = \kappa c(c_{k+1}-2c_k+c_{k-1}) + \kappa c_1(c_k-c_{k-1}) - \frac{1}{\tau} \, c_k \, , 
\label{ME_eq_k}
\ee
for $1<k<l$ and $l<k$,
\be
\frac{dc_1}{dt} = \kappa c(c_2-c_1) + \kappa c_1^2 + \frac{1}{\tau} (c_{1,0}-c_1) \, , \label{ME_eq_1}
\ee
and
\bea
\frac{dc_l}{dt} &=& \kappa c(c_{l+1}-2c_l+c_{l-1}) + \kappa c_1(c_l-c_{l-1}) \nonumber\\
&& \qquad\qquad\qquad\qquad\quad\ + \frac{1}{\tau} (c_{l,0}-c_1) \, , \label{ME_eq_l}
\eea
where $c=\sum_{k=1}^{\infty}c_k$ is the sum of all the concentrations.  Under the conditions of stationarity, this sum has reached its asymptotic value $c=c_{1,0}+c_{l,0}$, while the concentrations no longer depend on time: $dc_k/dt=0$.  Under such conditions, Eqs.~(\ref{ME_eq_k})-(\ref{ME_eq_l}) form a set of linear equations for the concentrations $\{c_k\}_{k=1}^{\infty}$.  The stationary solution is thus given by
\be
c_{k,{\rm st}} = \left\{
\begin{array}{ll}
A \, \Lambda_+^k + B \, \Lambda_-^k \, , &\qquad\mbox{for} \quad 1\leq k \leq l \, ,\\
C \, \Lambda_-^k \, , &\qquad\mbox{for} \quad l\leq k \, ,
\end{array}
\right.
\ee
in terms of the roots of the characteristic polynomial:
\be
\Lambda^2 + \left( \frac{c_1}{c}-2 -\frac{1}{c\kappa\tau}\right) \Lambda + 1 - \frac{c_1}{c} = 0
\label{charact_polyn}
\ee
with $\Lambda_-<1$.

If $\tau=\infty$, we recover the equilibrium exponential distribution~(\ref{ME_equil}) satisfying the conditions of detailed balance. 
In this case, the roots of Eq.~(\ref{charact_polyn}) are $\Lambda_+=1$ and $\Lambda_-=1-c_1/c$.  The normalizable distribution is thus given by $c_k=C\Lambda_-^k$ for $k \geq 1$, so that $C=cc_1/(c-c_1)$.  If $\tau=\infty$, the reactor is closed so that the quantities $c=\sum_{k=1}^{\infty}c_k$ and $M=\sum_{k=1}^{\infty}k c_k$ are invariant and they keep their initial values $c(0)=c$ and $M(0)=c^2/c_1$, hence the equilibrium distribution~(\ref{ME_equil}).

If $\tau$ is finite, the tail of the nonequilibrium distribution is still exponential, but the decay factor $\Lambda_-<1$ takes a different value than at equilibrium.  If $\tau$ is large enough, the decay factor is approximately given by
\be
\Lambda_- \simeq 1 -\frac{c_1(\infty)}{c(\infty)} + O(\tau^{-1}) \, ,
\ee
and $c_1(\infty)=c(\infty)^2/M(\infty)+O(\tau^{-1})$.
In this limit, the stationary distribution at large but finite values of the residence time is given by
\be
c_{k,{\rm st}} \simeq   \frac{c(\infty)^2}{M(\infty)} \left[ 1 - \frac{c(\infty)}{M(\infty)}\right]^{k-1} , \qquad\mbox{if}\quad c(\infty)\kappa\tau \gg 1 \, .
\label{ME_nonequil}
\ee
However, the sum of all the concentrations is no longer a conserved quantity in the open reactor where it converges towards its injection value: $c(\infty)=c_0=c_{1,0}+c_{l,0}$.  In the example of Fig.~\ref{fig2}, the injection values of the two conserved quantities are respectively equal to $c(\infty)=c_0=3$ and $M(\infty)=M_0=21$, although their initial values are $c(0)=0.75331$ and $M(0)=0.9119$.  This explains the observation in Fig.~\ref{fig2} that the distribution is decreasing more slowly as $c_{k,{\rm st}}\simeq 0.5\times(6/7)^k\simeq 0.5\times 0.857^k$ if $\tau=1000$ in the open reactor, than in the closed reactor if $\tau=\infty$.

If $\tau$ is small enough, the distribution becomes bimodal with two peaks at $c_{1,{\rm st}}\simeq c_{1,0}$ and $c_{l,{\rm st}}\simeq c_{l,0}$.  In this limit, the tail of the distribution behaves as $c_k\simeq c_{l,0}\Lambda_-^{k-l}$ for $k\geq l$ with $\Lambda_-\simeq c_{l,0}\kappa\tau +O[(\kappa\tau)^2]$, so that the decay can be faster than at equilibrium, as seen in Fig.~\ref{fig2}.


\vskip 0.5 cm


\begin{thebibliography}{0}%
\makeatletter
\providecommand \@ifxundefined [1]{%
 \@ifx{#1\undefined}
}%
\providecommand \@ifnum [1]{%
 \ifnum #1\expandafter \@firstoftwo
 \else \expandafter \@secondoftwo
 \fi
}%
\providecommand \@ifx [1]{%
 \ifx #1\expandafter \@firstoftwo
 \else \expandafter \@secondoftwo
 \fi
}%
\providecommand \natexlab [1]{#1}%
\providecommand \enquote  [1]{``#1''}%
\providecommand \bibnamefont  [1]{#1}%
\providecommand \bibfnamefont [1]{#1}%
\providecommand \citenamefont [1]{#1}%
\providecommand \href@noop [0]{\@secondoftwo}%
\providecommand \href [0]{\begingroup \@sanitize@url \@href}%
\providecommand \@href[1]{\@@startlink{#1}\@@href}%
\providecommand \@@href[1]{\endgroup#1\@@endlink}%
\providecommand \@sanitize@url [0]{\catcode `\\12\catcode `\$12\catcode
  `\&12\catcode `\#12\catcode `\^12\catcode `\_12\catcode `\%12\relax}%
\providecommand \@@startlink[1]{}%
\providecommand \@@endlink[0]{}%
\providecommand \url  [0]{\begingroup\@sanitize@url \@url }%
\providecommand \@url [1]{\endgroup\@href {#1}{\urlprefix }}%
\providecommand \urlprefix  [0]{URL }%
\providecommand \Eprint [0]{\href }%
\providecommand \doibase [0]{http://dx.doi.org/}%
\providecommand \selectlanguage [0]{\@gobble}%
\providecommand \bibinfo  [0]{\@secondoftwo}%
\providecommand \bibfield  [0]{\@secondoftwo}%
\providecommand \translation [1]{[#1]}%
\providecommand \BibitemOpen [0]{}%
\providecommand \bibitemStop [0]{}%
\providecommand \bibitemNoStop [0]{.\EOS\space}%
\providecommand \EOS [0]{\spacefactor3000\relax}%
\providecommand \BibitemShut  [1]{\csname bibitem#1\endcsname}%
\let\auto@bib@innerbib\@empty
\end{thebibliography}%


\begin{thebibliography}{38}%
\makeatletter
\providecommand \@ifxundefined [1]{%
 \@ifx{#1\undefined}
}%
\providecommand \@ifnum [1]{%
 \ifnum #1\expandafter \@firstoftwo
 \else \expandafter \@secondoftwo
 \fi
}%
\providecommand \@ifx [1]{%
 \ifx #1\expandafter \@firstoftwo
 \else \expandafter \@secondoftwo
 \fi
}%
\providecommand \natexlab [1]{#1}%
\providecommand \enquote  [1]{``#1''}%
\providecommand \bibnamefont  [1]{#1}%
\providecommand \bibfnamefont [1]{#1}%
\providecommand \citenamefont [1]{#1}%
\providecommand \href@noop [0]{\@secondoftwo}%
\providecommand \href [0]{\begingroup \@sanitize@url \@href}%
\providecommand \@href[1]{\@@startlink{#1}\@@href}%
\providecommand \@@href[1]{\endgroup#1\@@endlink}%
\providecommand \@sanitize@url [0]{\catcode `\\12\catcode `\$12\catcode
  `\&12\catcode `\#12\catcode `\^12\catcode `\_12\catcode `\%12\relax}%
\providecommand \@@startlink[1]{}%
\providecommand \@@endlink[0]{}%
\providecommand \url  [0]{\begingroup\@sanitize@url \@url }%
\providecommand \@url [1]{\endgroup\@href {#1}{\urlprefix }}%
\providecommand \urlprefix  [0]{URL }%
\providecommand \Eprint [0]{\href }%
\providecommand \doibase [0]{http://dx.doi.org/}%
\providecommand \selectlanguage [0]{\@gobble}%
\providecommand \bibinfo  [0]{\@secondoftwo}%
\providecommand \bibfield  [0]{\@secondoftwo}%
\providecommand \translation [1]{[#1]}%
\providecommand \BibitemOpen [0]{}%
\providecommand \bibitemStop [0]{}%
\providecommand \bibitemNoStop [0]{.\EOS\space}%
\providecommand \EOS [0]{\spacefactor3000\relax}%
\providecommand \BibitemShut  [1]{\csname bibitem#1\endcsname}%
\let\auto@bib@innerbib\@empty
\bibitem [{\citenamefont {Vaidya}\ \emph {et~al.}(2012)\citenamefont {Vaidya},
  \citenamefont {Manapat}, \citenamefont {Chen}, \citenamefont {Xulvi-Brunet},
  \citenamefont {Hayden},\ and\ \citenamefont {Lehman}}]{Vaidya2012}%
  \BibitemOpen
  \bibfield  {author} {\bibinfo {author} {\bibfnamefont {N.}~\bibnamefont
  {Vaidya}}, \bibinfo {author} {\bibfnamefont {M.~L.}\ \bibnamefont {Manapat}},
  \bibinfo {author} {\bibfnamefont {I.~A.}\ \bibnamefont {Chen}}, \bibinfo
  {author} {\bibfnamefont {R.}~\bibnamefont {Xulvi-Brunet}}, \bibinfo {author}
  {\bibfnamefont {E.~J.}\ \bibnamefont {Hayden}}, \ and\ \bibinfo {author}
  {\bibfnamefont {N.}~\bibnamefont {Lehman}},\ }\href@noop {} {\bibfield
  {journal} {\bibinfo  {journal} {Nature}\ }\textbf {\bibinfo {volume} {491}},\
  \bibinfo {pages} {72} (\bibinfo {year} {2012})}\BibitemShut {NoStop}%
\bibitem [{\citenamefont {Forsythe}\ \emph {et~al.}(2017)\citenamefont
  {Forsythe}, \citenamefont {Petrov}, \citenamefont {Millar}, \citenamefont
  {Yu}, \citenamefont {Krishnamurthy}, \citenamefont {Grover}, \citenamefont
  {Hud},\ and\ \citenamefont {Fernandez}}]{Forsythe2017}%
  \BibitemOpen
  \bibfield  {author} {\bibinfo {author} {\bibfnamefont {J.~G.}\ \bibnamefont
  {Forsythe}}, \bibinfo {author} {\bibfnamefont {A.~S.}\ \bibnamefont
  {Petrov}}, \bibinfo {author} {\bibfnamefont {W.~C.}\ \bibnamefont {Millar}},
  \bibinfo {author} {\bibfnamefont {S.-S.}\ \bibnamefont {Yu}}, \bibinfo
  {author} {\bibfnamefont {R.}~\bibnamefont {Krishnamurthy}}, \bibinfo {author}
  {\bibfnamefont {M.~A.}\ \bibnamefont {Grover}}, \bibinfo {author}
  {\bibfnamefont {N.~V.}\ \bibnamefont {Hud}}, \ and\ \bibinfo {author}
  {\bibfnamefont {F.~M.}\ \bibnamefont {Fernandez}},\ }\href {\doibase
  10.1073/pnas.1711631114} {\bibfield  {journal} {\bibinfo  {journal} {Proc.
  Natl. Acad. Sci. U.S.A.}\ }\textbf {\bibinfo {volume} {114}},\ \bibinfo
  {pages} {E7652} (\bibinfo {year} {2017})}\BibitemShut {NoStop}%
\bibitem [{\citenamefont {Eigen}(1971)}]{Eigen1971}%
  \BibitemOpen
  \bibfield  {author} {\bibinfo {author} {\bibfnamefont {M.}~\bibnamefont
  {Eigen}},\ }\href@noop {} {\bibfield  {journal} {\bibinfo  {journal}
  {Naturwissenschaften}\ }\textbf {\bibinfo {volume} {58}},\ \bibinfo {pages}
  {465} (\bibinfo {year} {1971})}\BibitemShut {NoStop}%
\bibitem [{\citenamefont {Eigen}\ and\ \citenamefont {Schuster}(1977)}]{EP77}%
  \BibitemOpen
  \bibfield  {author} {\bibinfo {author} {\bibfnamefont {M.}~\bibnamefont
  {Eigen}}\ and\ \bibinfo {author} {\bibfnamefont {P.}~\bibnamefont
  {Schuster}},\ }\href@noop {} {\bibfield  {journal} {\bibinfo  {journal}
  {Naturwissenschaften}\ }\textbf {\bibinfo {volume} {64}},\ \bibinfo {pages}
  {541} (\bibinfo {year} {1977})}\BibitemShut {NoStop}%
\bibitem [{\citenamefont {Eigen}\ and\ \citenamefont
  {Schuster}(1978{\natexlab{a}})}]{EP78a}%
  \BibitemOpen
  \bibfield  {author} {\bibinfo {author} {\bibfnamefont {M.}~\bibnamefont
  {Eigen}}\ and\ \bibinfo {author} {\bibfnamefont {P.}~\bibnamefont
  {Schuster}},\ }\href@noop {} {\bibfield  {journal} {\bibinfo  {journal}
  {Naturwissenschaften}\ }\textbf {\bibinfo {volume} {65}},\ \bibinfo {pages}
  {7} (\bibinfo {year} {1978}{\natexlab{a}})}\BibitemShut {NoStop}%
\bibitem [{\citenamefont {Eigen}\ and\ \citenamefont
  {Schuster}(1978{\natexlab{b}})}]{EP78b}%
  \BibitemOpen
  \bibfield  {author} {\bibinfo {author} {\bibfnamefont {M.}~\bibnamefont
  {Eigen}}\ and\ \bibinfo {author} {\bibfnamefont {P.}~\bibnamefont
  {Schuster}},\ }\href@noop {} {\bibfield  {journal} {\bibinfo  {journal}
  {Naturwissenschaften}\ }\textbf {\bibinfo {volume} {65}},\ \bibinfo {pages}
  {341} (\bibinfo {year} {1978}{\natexlab{b}})}\BibitemShut {NoStop}%
\bibitem [{\citenamefont {Eigen}(1993)}]{Eigen1992}%
  \BibitemOpen
  \bibfield  {author} {\bibinfo {author} {\bibfnamefont {M.}~\bibnamefont
  {Eigen}},\ }\href@noop {} {\emph {\bibinfo {title} {Steps towards Life: A
  Perspective on Evolution}}}\ (\bibinfo  {publisher} {Oxford University
  Press},\ \bibinfo {address} {Oxford},\ \bibinfo {year} {1993})\BibitemShut
  {NoStop}%
\bibitem [{\citenamefont {Kauffman}(1993)}]{Kauffman1993}%
  \BibitemOpen
  \bibfield  {author} {\bibinfo {author} {\bibfnamefont {S.~A.}\ \bibnamefont
  {Kauffman}},\ }\href@noop {} {\emph {\bibinfo {title} {The Origins of Order:
  Self-Organization and Selection in Evolution}}},\ edited by\ \bibinfo
  {editor} {\bibfnamefont {O.~U.~P.}\ \bibnamefont {Inc.}}\ (\bibinfo
  {publisher} {Oxford University Press},\ \bibinfo {year} {1993})\BibitemShut
  {NoStop}%
\bibitem [{\citenamefont {Segr{\'e}}\ \emph {et~al.}(2000)\citenamefont
  {Segr{\'e}}, \citenamefont {Ben-Eli},\ and\ \citenamefont
  {Lancet}}]{Segre2000}%
  \BibitemOpen
  \bibfield  {author} {\bibinfo {author} {\bibfnamefont {D.}~\bibnamefont
  {Segr{\'e}}}, \bibinfo {author} {\bibfnamefont {D.}~\bibnamefont {Ben-Eli}},
  \ and\ \bibinfo {author} {\bibfnamefont {D.}~\bibnamefont {Lancet}},\ }\href
  {\doibase 10.1073/pnas.97.8.4112} {\bibfield  {journal} {\bibinfo  {journal}
  {Proc. Natl. Acad. Sci. U.S.A.}\ }\textbf {\bibinfo {volume} {97}},\ \bibinfo
  {pages} {4112} (\bibinfo {year} {2000})}\BibitemShut {NoStop}%
\bibitem [{\citenamefont {Lincoln}\ and\ \citenamefont
  {Joyce}(2009)}]{Lincoln2009}%
  \BibitemOpen
  \bibfield  {author} {\bibinfo {author} {\bibfnamefont {T.~A.}\ \bibnamefont
  {Lincoln}}\ and\ \bibinfo {author} {\bibfnamefont {G.~F.}\ \bibnamefont
  {Joyce}},\ }\href {\doibase 10.1126/science.1167856} {\bibfield  {journal}
  {\bibinfo  {journal} {Science}\ }\textbf {\bibinfo {volume} {323}},\ \bibinfo
  {pages} {1229} (\bibinfo {year} {2009})}\BibitemShut {NoStop}%
\bibitem [{\citenamefont {Matsumura}\ \emph {et~al.}(2016)\citenamefont
  {Matsumura}, \citenamefont {Kun}, \citenamefont {Ryckelynck}, \citenamefont
  {Coldren}, \citenamefont {Szil{\'a}gyi}, \citenamefont {Jossinet},
  \citenamefont {Rick}, \citenamefont {Nghe}, \citenamefont {Szathm{\'a}ry},\
  and\ \citenamefont {Griffiths}}]{Matsumura2016}%
  \BibitemOpen
  \bibfield  {author} {\bibinfo {author} {\bibfnamefont {S.}~\bibnamefont
  {Matsumura}}, \bibinfo {author} {\bibfnamefont {{\'A}.}~\bibnamefont {Kun}},
  \bibinfo {author} {\bibfnamefont {M.}~\bibnamefont {Ryckelynck}}, \bibinfo
  {author} {\bibfnamefont {F.}~\bibnamefont {Coldren}}, \bibinfo {author}
  {\bibfnamefont {A.}~\bibnamefont {Szil{\'a}gyi}}, \bibinfo {author}
  {\bibfnamefont {F.}~\bibnamefont {Jossinet}}, \bibinfo {author}
  {\bibfnamefont {C.}~\bibnamefont {Rick}}, \bibinfo {author} {\bibfnamefont
  {P.}~\bibnamefont {Nghe}}, \bibinfo {author} {\bibfnamefont {E.}~\bibnamefont
  {Szathm{\'a}ry}}, \ and\ \bibinfo {author} {\bibfnamefont {A.~D.}\
  \bibnamefont {Griffiths}},\ }\href {\doibase 10.1126/science.aag1582}
  {\bibfield  {journal} {\bibinfo  {journal} {Science}\ }\textbf {\bibinfo
  {volume} {354}},\ \bibinfo {pages} {1293} (\bibinfo {year}
  {2016})}\BibitemShut {NoStop}%
\bibitem [{\citenamefont {Zwaag}\ and\ \citenamefont
  {Meijer}(2015)}]{Zwaag2015}%
  \BibitemOpen
  \bibfield  {author} {\bibinfo {author} {\bibfnamefont {D.~v.~d.}\
  \bibnamefont {Zwaag}}\ and\ \bibinfo {author} {\bibfnamefont {E.~W.}\
  \bibnamefont {Meijer}},\ }\href {\doibase 10.1126/science.aad0194} {\bibfield
   {journal} {\bibinfo  {journal} {Science}\ }\textbf {\bibinfo {volume}
  {349}},\ \bibinfo {pages} {1056} (\bibinfo {year} {2015})}\BibitemShut
  {NoStop}%
\bibitem [{\citenamefont {Zeravcic}\ \emph {et~al.}(2017)\citenamefont
  {Zeravcic}, \citenamefont {Manoharan},\ and\ \citenamefont
  {Brenner}}]{Zeravcic2017}%
  \BibitemOpen
  \bibfield  {author} {\bibinfo {author} {\bibfnamefont {Z.}~\bibnamefont
  {Zeravcic}}, \bibinfo {author} {\bibfnamefont {V.~N.}\ \bibnamefont
  {Manoharan}}, \ and\ \bibinfo {author} {\bibfnamefont {M.~P.}\ \bibnamefont
  {Brenner}},\ }\href {\doibase 10.1103/RevModPhys.89.031001} {\bibfield
  {journal} {\bibinfo  {journal} {Rev. Mod. Phys.}\ }\textbf {\bibinfo {volume}
  {89}},\ \bibinfo {pages} {031001} (\bibinfo {year} {2017})}\BibitemShut
  {NoStop}%
\bibitem [{\citenamefont {Agresti}\ \emph {et~al.}(2010)\citenamefont
  {Agresti}, \citenamefont {Antipov}, \citenamefont {Abate}, \citenamefont
  {Ahn}, \citenamefont {Rowat}, \citenamefont {Baret}, \citenamefont {Marquez},
  \citenamefont {Klibanov}, \citenamefont {Griffiths},\ and\ \citenamefont
  {Weitz}}]{Agresti2010}%
  \BibitemOpen
  \bibfield  {author} {\bibinfo {author} {\bibfnamefont {J.~J.}\ \bibnamefont
  {Agresti}}, \bibinfo {author} {\bibfnamefont {E.}~\bibnamefont {Antipov}},
  \bibinfo {author} {\bibfnamefont {A.~R.}\ \bibnamefont {Abate}}, \bibinfo
  {author} {\bibfnamefont {K.}~\bibnamefont {Ahn}}, \bibinfo {author}
  {\bibfnamefont {A.~C.}\ \bibnamefont {Rowat}}, \bibinfo {author}
  {\bibfnamefont {J.-C.}\ \bibnamefont {Baret}}, \bibinfo {author}
  {\bibfnamefont {M.}~\bibnamefont {Marquez}}, \bibinfo {author} {\bibfnamefont
  {A.~M.}\ \bibnamefont {Klibanov}}, \bibinfo {author} {\bibfnamefont {A.~D.}\
  \bibnamefont {Griffiths}}, \ and\ \bibinfo {author} {\bibfnamefont {D.~A.}\
  \bibnamefont {Weitz}},\ }\href {\doibase 10.1073/pnas.0910781107} {\bibfield
  {journal} {\bibinfo  {journal} {Proc. Natl. Acad. Sci. U.S.A.}\ }\textbf
  {\bibinfo {volume} {107}},\ \bibinfo {pages} {4004} (\bibinfo {year}
  {2010})}\BibitemShut {NoStop}%
\bibitem [{\citenamefont {Arnold}\ and\ \citenamefont
  {Moore}(1997)}]{Arnold1997}%
  \BibitemOpen
  \bibfield  {author} {\bibinfo {author} {\bibfnamefont {F.}~\bibnamefont
  {Arnold}}\ and\ \bibinfo {author} {\bibfnamefont {J.}~\bibnamefont {Moore}},\
  }\href@noop {} {\bibfield  {journal} {\bibinfo  {journal} {Adv. Biochem. Eng.
  Biotechnol.}\ }\textbf {\bibinfo {volume} {58}},\ \bibinfo {pages} {1}
  (\bibinfo {year} {1997})}\BibitemShut {NoStop}%
\bibitem [{\citenamefont {Hill}(1989)}]{Hill1989}%
  \BibitemOpen
  \bibfield  {author} {\bibinfo {author} {\bibfnamefont {T.~L.}\ \bibnamefont
  {Hill}},\ }\href@noop {} {\emph {\bibinfo {title} {Free Energy Transduction
  and Biochemical Cycle Kinetics}}}\ (\bibinfo  {publisher} {Springer},\
  \bibinfo {year} {1989})\BibitemShut {NoStop}%
\bibitem [{\citenamefont {Padinhateeri}\ \emph {et~al.}(2012)\citenamefont
  {Padinhateeri}, \citenamefont {Kolomeisky},\ and\ \citenamefont
  {Lacoste}}]{Ranjith2012}%
  \BibitemOpen
  \bibfield  {author} {\bibinfo {author} {\bibfnamefont {R.}~\bibnamefont
  {Padinhateeri}}, \bibinfo {author} {\bibfnamefont {A.}~\bibnamefont
  {Kolomeisky}}, \ and\ \bibinfo {author} {\bibfnamefont {D.}~\bibnamefont
  {Lacoste}},\ }\href {\doibase 10.1016/j.bpj.2011.12.059} {\bibfield
  {journal} {\bibinfo  {journal} {Biophys. J.}\ }\textbf {\bibinfo {volume}
  {102}},\ \bibinfo {pages} {1274} (\bibinfo {year} {2012})}\BibitemShut
  {NoStop}%
\bibitem [{\citenamefont {J{\'e}gou}\ and\ \citenamefont
  {Romet-Lemonne}(2016)}]{Jegou2016}%
  \BibitemOpen
  \bibfield  {author} {\bibinfo {author} {\bibfnamefont {A.}~\bibnamefont
  {J{\'e}gou}}\ and\ \bibinfo {author} {\bibfnamefont {G.}~\bibnamefont
  {Romet-Lemonne}},\ }\bibfield  {booktitle} {\emph {\bibinfo {booktitle}
  {Biophysical Journal}},\ }\href {\doibase 10.1016/j.bpj.2016.04.025}
  {\bibfield  {journal} {\bibinfo  {journal} {Biophys. J.}\ }\textbf {\bibinfo
  {volume} {110}},\ \bibinfo {pages} {2138} (\bibinfo {year}
  {2016})}\BibitemShut {NoStop}%
\bibitem [{\citenamefont {Brangwynne}\ \emph {et~al.}(2009)\citenamefont
  {Brangwynne}, \citenamefont {Eckmann}, \citenamefont {Courson}, \citenamefont
  {Rybarska}, \citenamefont {Hoege}, \citenamefont {Gharakhani}, \citenamefont
  {J{\"u}licher},\ and\ \citenamefont {Hyman}}]{Brangwynne2009}%
  \BibitemOpen
  \bibfield  {author} {\bibinfo {author} {\bibfnamefont {C.~P.}\ \bibnamefont
  {Brangwynne}}, \bibinfo {author} {\bibfnamefont {C.~R.}\ \bibnamefont
  {Eckmann}}, \bibinfo {author} {\bibfnamefont {D.~S.}\ \bibnamefont
  {Courson}}, \bibinfo {author} {\bibfnamefont {A.}~\bibnamefont {Rybarska}},
  \bibinfo {author} {\bibfnamefont {C.}~\bibnamefont {Hoege}}, \bibinfo
  {author} {\bibfnamefont {J.}~\bibnamefont {Gharakhani}}, \bibinfo {author}
  {\bibfnamefont {F.}~\bibnamefont {J{\"u}licher}}, \ and\ \bibinfo {author}
  {\bibfnamefont {A.~A.}\ \bibnamefont {Hyman}},\ }\href@noop {} {\bibfield
  {journal} {\bibinfo  {journal} {Science}\ }\textbf {\bibinfo {volume}
  {324}},\ \bibinfo {pages} {1729} (\bibinfo {year} {2009})}\BibitemShut
  {NoStop}%
\bibitem [{\citenamefont {Zwicker}\ \emph {et~al.}(2017)\citenamefont
  {Zwicker}, \citenamefont {Seyboldt}, \citenamefont {Weber}, \citenamefont
  {Hyman},\ and\ \citenamefont {J{\"u}licher}}]{Zwicker2017}%
  \BibitemOpen
  \bibfield  {author} {\bibinfo {author} {\bibfnamefont {D.}~\bibnamefont
  {Zwicker}}, \bibinfo {author} {\bibfnamefont {R.}~\bibnamefont {Seyboldt}},
  \bibinfo {author} {\bibfnamefont {C.~A.}\ \bibnamefont {Weber}}, \bibinfo
  {author} {\bibfnamefont {A.~A.}\ \bibnamefont {Hyman}}, \ and\ \bibinfo
  {author} {\bibfnamefont {F.}~\bibnamefont {J{\"u}licher}},\ }\href@noop {}
  {\bibfield  {journal} {\bibinfo  {journal} {Nat. Phys.}\ }\textbf {\bibinfo
  {volume} {13}},\ \bibinfo {pages} {408} (\bibinfo {year} {2017})}\BibitemShut
  {NoStop}%
\bibitem [{\citenamefont {Marchetti}\ \emph {et~al.}(2013)\citenamefont
  {Marchetti}, \citenamefont {Joanny}, \citenamefont {Ramaswamy}, \citenamefont
  {Liverpool}, \citenamefont {Prost}, \citenamefont {Rao},\ and\ \citenamefont
  {Simha}}]{Marchetti2013}%
  \BibitemOpen
  \bibfield  {author} {\bibinfo {author} {\bibfnamefont {M.~C.}\ \bibnamefont
  {Marchetti}}, \bibinfo {author} {\bibfnamefont {J.~F.}\ \bibnamefont
  {Joanny}}, \bibinfo {author} {\bibfnamefont {S.}~\bibnamefont {Ramaswamy}},
  \bibinfo {author} {\bibfnamefont {T.~B.}\ \bibnamefont {Liverpool}}, \bibinfo
  {author} {\bibfnamefont {J.}~\bibnamefont {Prost}}, \bibinfo {author}
  {\bibfnamefont {M.}~\bibnamefont {Rao}}, \ and\ \bibinfo {author}
  {\bibfnamefont {R.~A.}\ \bibnamefont {Simha}},\ }\href {\doibase
  10.1103/RevModPhys.85.1143} {\bibfield  {journal} {\bibinfo  {journal} {Rev.
  Mod. Phys.}\ }\textbf {\bibinfo {volume} {85}},\ \bibinfo {pages} {1143}
  (\bibinfo {year} {2013})}\BibitemShut {NoStop}%
\bibitem [{\citenamefont {Nicolis}\ and\ \citenamefont
  {Prigogine}(1977)}]{NP77}%
  \BibitemOpen
  \bibfield  {author} {\bibinfo {author} {\bibfnamefont {G.}~\bibnamefont
  {Nicolis}}\ and\ \bibinfo {author} {\bibfnamefont {I.}~\bibnamefont
  {Prigogine}},\ }\href@noop {} {\emph {\bibinfo {title} {Self-Organization in
  Nonequilibrium Systems: From Dissipative Structures to Order through
  Fluctuations}}}\ (\bibinfo  {publisher} {Wiley},\ \bibinfo {address} {New
  York},\ \bibinfo {year} {1977})\BibitemShut {NoStop}%
\bibitem [{\citenamefont {Prigogine}(1967)}]{P67}%
  \BibitemOpen
  \bibfield  {author} {\bibinfo {author} {\bibfnamefont {I.}~\bibnamefont
  {Prigogine}},\ }\href@noop {} {\emph {\bibinfo {title} {Introduction to
  Thermodynamics of Irreversible Processes}}}\ (\bibinfo  {publisher} {Wiley},\
  \bibinfo {address} {New York},\ \bibinfo {year} {1967})\BibitemShut {NoStop}%
\bibitem [{\citenamefont {de~Groot}\ and\ \citenamefont {Mazur}(1984)}]{GM84}%
  \BibitemOpen
  \bibfield  {author} {\bibinfo {author} {\bibfnamefont {S.~R.}\ \bibnamefont
  {de~Groot}}\ and\ \bibinfo {author} {\bibfnamefont {P.}~\bibnamefont
  {Mazur}},\ }\href@noop {} {\emph {\bibinfo {title} {Nonequilibrium
  Thermodynamics}}}\ (\bibinfo  {publisher} {Dover},\ \bibinfo {address} {New
  York},\ \bibinfo {year} {1984})\BibitemShut {NoStop}%
\bibitem [{\citenamefont {Polettini}\ and\ \citenamefont
  {Esposito}(2014)}]{Polettini2014}%
  \BibitemOpen
  \bibfield  {author} {\bibinfo {author} {\bibfnamefont {M.}~\bibnamefont
  {Polettini}}\ and\ \bibinfo {author} {\bibfnamefont {M.}~\bibnamefont
  {Esposito}},\ }\href {\doibase http://dx.doi.org/10.1063/1.4886396}
  {\bibfield  {journal} {\bibinfo  {journal} {J. Chem. Phys.}\ }\textbf
  {\bibinfo {volume} {141}},\ \bibinfo {eid} {024117} (\bibinfo {year}
  {2014})}\BibitemShut {NoStop}%
\bibitem [{\citenamefont {Rao}\ and\ \citenamefont {Esposito}(2016)}]{RE16}%
  \BibitemOpen
  \bibfield  {author} {\bibinfo {author} {\bibfnamefont {R.}~\bibnamefont
  {Rao}}\ and\ \bibinfo {author} {\bibfnamefont {M.}~\bibnamefont {Esposito}},\
  }\href@noop {} {\bibfield  {journal} {\bibinfo  {journal} {Phys. Rev. X}\
  }\textbf {\bibinfo {volume} {6}},\ \bibinfo {pages} {041064} (\bibinfo {year}
  {2016})}\BibitemShut {NoStop}%
\bibitem [{\citenamefont {Lahiri}\ \emph {et~al.}(2015)\citenamefont {Lahiri},
  \citenamefont {Wang}, \citenamefont {Esposito},\ and\ \citenamefont
  {Lacoste}}]{Lahiri2015}%
  \BibitemOpen
  \bibfield  {author} {\bibinfo {author} {\bibfnamefont {S.}~\bibnamefont
  {Lahiri}}, \bibinfo {author} {\bibfnamefont {Y.}~\bibnamefont {Wang}},
  \bibinfo {author} {\bibfnamefont {M.}~\bibnamefont {Esposito}}, \ and\
  \bibinfo {author} {\bibfnamefont {D.}~\bibnamefont {Lacoste}},\ }\href@noop
  {} {\bibfield  {journal} {\bibinfo  {journal} {New J. Phys.}\ }\textbf
  {\bibinfo {volume} {17}},\ \bibinfo {pages} {085008} (\bibinfo {year}
  {2015})}\BibitemShut {NoStop}%
\bibitem [{\citenamefont {Rao}\ \emph {et~al.}(2015)\citenamefont {Rao},
  \citenamefont {Lacoste},\ and\ \citenamefont {Esposito}}]{Rao2015a}%
  \BibitemOpen
  \bibfield  {author} {\bibinfo {author} {\bibfnamefont {R.}~\bibnamefont
  {Rao}}, \bibinfo {author} {\bibfnamefont {D.}~\bibnamefont {Lacoste}}, \ and\
  \bibinfo {author} {\bibfnamefont {M.}~\bibnamefont {Esposito}},\ }\href@noop
  {} {\bibfield  {journal} {\bibinfo  {journal} {J. Chem. Phys.}\ }\textbf
  {\bibinfo {volume} {143}},\ \bibinfo {pages} {244903} (\bibinfo {year}
  {2015})}\BibitemShut {NoStop}%
\bibitem [{\citenamefont {Blokhuis}\ and\ \citenamefont
  {Lacoste}(2017)}]{Blokhuis2017}%
  \BibitemOpen
  \bibfield  {author} {\bibinfo {author} {\bibfnamefont {A.}~\bibnamefont
  {Blokhuis}}\ and\ \bibinfo {author} {\bibfnamefont {D.}~\bibnamefont
  {Lacoste}},\ }\href {\doibase 10.1063/1.5001021} {\bibfield  {journal}
  {\bibinfo  {journal} {J. Chem. Phys.}\ }\textbf {\bibinfo {volume} {147}},\
  \bibinfo {pages} {094905} (\bibinfo {year} {2017})}\BibitemShut {NoStop}%
\bibitem [{\citenamefont {Aris}(1989)}]{A89}%
  \BibitemOpen
  \bibfield  {author} {\bibinfo {author} {\bibfnamefont {R.}~\bibnamefont
  {Aris}},\ }\href@noop {} {\emph {\bibinfo {title} {Elementary Chemical
  Reactor Analysis}}}\ (\bibinfo  {publisher} {Dover},\ \bibinfo {address}
  {Mineola NY},\ \bibinfo {year} {1989})\BibitemShut {NoStop}%
\bibitem [{\citenamefont {Berg\'e}\ \emph {et~al.}(1984)\citenamefont
  {Berg\'e}, \citenamefont {Pomeau},\ and\ \citenamefont {Vidal}}]{BPV84}%
  \BibitemOpen
  \bibfield  {author} {\bibinfo {author} {\bibfnamefont {P.}~\bibnamefont
  {Berg\'e}}, \bibinfo {author} {\bibfnamefont {Y.}~\bibnamefont {Pomeau}}, \
  and\ \bibinfo {author} {\bibfnamefont {C.}~\bibnamefont {Vidal}},\
  }\href@noop {} {\emph {\bibinfo {title} {L'ordre dans le chaos}}}\ (\bibinfo
  {publisher} {Hermann},\ \bibinfo {address} {Paris},\ \bibinfo {year}
  {1984})\BibitemShut {NoStop}%
\bibitem [{\citenamefont {Vidal}\ and\ \citenamefont
  {Lemarchand}(1988)}]{VL88}%
  \BibitemOpen
  \bibfield  {author} {\bibinfo {author} {\bibfnamefont {C.}~\bibnamefont
  {Vidal}}\ and\ \bibinfo {author} {\bibfnamefont {H.}~\bibnamefont
  {Lemarchand}},\ }\href@noop {} {\emph {\bibinfo {title} {La r\'eaction
  cr\'eatrice: Dynamique des syst\`emes chimiques}}}\ (\bibinfo  {publisher}
  {Hermann},\ \bibinfo {address} {Paris},\ \bibinfo {year} {1988})\BibitemShut
  {NoStop}%
\bibitem [{\citenamefont {Nicolis}(1995)}]{N95}%
  \BibitemOpen
  \bibfield  {author} {\bibinfo {author} {\bibfnamefont {G.}~\bibnamefont
  {Nicolis}},\ }\href@noop {} {\emph {\bibinfo {title} {Introduction to
  nonlinear science}}}\ (\bibinfo  {publisher} {Cambridge University Press},\
  \bibinfo {address} {Cambridge UK},\ \bibinfo {year} {1995})\BibitemShut
  {NoStop}%
\bibitem [{\citenamefont {Epstein}\ and\ \citenamefont {Pojman}(1998)}]{EP98}%
  \BibitemOpen
  \bibfield  {author} {\bibinfo {author} {\bibfnamefont {I.~R.}\ \bibnamefont
  {Epstein}}\ and\ \bibinfo {author} {\bibfnamefont {J.~A.}\ \bibnamefont
  {Pojman}},\ }\href@noop {} {\emph {\bibinfo {title} {An Introduction to
  Nonlinear Chemical Dynamics}}}\ (\bibinfo  {publisher} {Oxford University
  Press},\ \bibinfo {address} {New York},\ \bibinfo {year} {1998})\BibitemShut
  {NoStop}%
\bibitem [{\citenamefont {Scott}(1991)}]{S91}%
  \BibitemOpen
  \bibfield  {author} {\bibinfo {author} {\bibfnamefont {S.~K.}\ \bibnamefont
  {Scott}},\ }\href@noop {} {\emph {\bibinfo {title} {Chemical Chaos}}}\
  (\bibinfo  {publisher} {Clarendon Press},\ \bibinfo {address} {Oxford},\
  \bibinfo {year} {1991})\BibitemShut {NoStop}%
\bibitem [{\citenamefont {Salman}\ \emph {et~al.}(2012)\citenamefont {Salman},
  \citenamefont {Brenner}, \citenamefont {Tung}, \citenamefont {Elyahu},
  \citenamefont {Stolovicki}, \citenamefont {Moore}, \citenamefont
  {Libchaber},\ and\ \citenamefont {Braun}}]{Salman2012}%
  \BibitemOpen
  \bibfield  {author} {\bibinfo {author} {\bibfnamefont {H.}~\bibnamefont
  {Salman}}, \bibinfo {author} {\bibfnamefont {N.}~\bibnamefont {Brenner}},
  \bibinfo {author} {\bibfnamefont {C.-k.}\ \bibnamefont {Tung}}, \bibinfo
  {author} {\bibfnamefont {N.}~\bibnamefont {Elyahu}}, \bibinfo {author}
  {\bibfnamefont {E.}~\bibnamefont {Stolovicki}}, \bibinfo {author}
  {\bibfnamefont {L.}~\bibnamefont {Moore}}, \bibinfo {author} {\bibfnamefont
  {A.}~\bibnamefont {Libchaber}}, \ and\ \bibinfo {author} {\bibfnamefont
  {E.}~\bibnamefont {Braun}},\ }\href {\doibase 10.1103/PhysRevLett.108.238105}
  {\bibfield  {journal} {\bibinfo  {journal} {Phys. Rev. Lett.}\ }\textbf
  {\bibinfo {volume} {108}},\ \bibinfo {pages} {238105} (\bibinfo {year}
  {2012})}\BibitemShut {NoStop}%
\bibitem [{\citenamefont {Tkachenko}\ and\ \citenamefont
  {Maslov}(2015)}]{Tkachenko2015}%
  \BibitemOpen
  \bibfield  {author} {\bibinfo {author} {\bibfnamefont {A.~V.}\ \bibnamefont
  {Tkachenko}}\ and\ \bibinfo {author} {\bibfnamefont {S.}~\bibnamefont
  {Maslov}},\ }\href {\doibase 10.1063/1.4922545} {\bibfield  {journal}
  {\bibinfo  {journal} {J. Chem. Phys.}\ }\textbf {\bibinfo {volume} {143}},\
  \bibinfo {pages} {045102} (\bibinfo {year} {2015})}\BibitemShut {NoStop}%
\bibitem [{\citenamefont {Mast}\ \emph {et~al.}(2013)\citenamefont {Mast},
  \citenamefont {Schink}, \citenamefont {Gerland},\ and\ \citenamefont
  {Braun}}]{Mast2013}%
  \BibitemOpen
  \bibfield  {author} {\bibinfo {author} {\bibfnamefont {C.~B.}\ \bibnamefont
  {Mast}}, \bibinfo {author} {\bibfnamefont {S.}~\bibnamefont {Schink}},
  \bibinfo {author} {\bibfnamefont {U.}~\bibnamefont {Gerland}}, \ and\
  \bibinfo {author} {\bibfnamefont {D.}~\bibnamefont {Braun}},\ }\href
  {\doibase 10.1073/pnas.1303222110} {\bibfield  {journal} {\bibinfo  {journal}
  {Proc. Natl. Acad. Sci. U.S.A.}\ }\textbf {\bibinfo {volume} {110}},\
  \bibinfo {pages} {8030} (\bibinfo {year} {2013})}\BibitemShut {NoStop}%
\end{thebibliography}

\end{document}